\documentclass[AMA,Times1COL]{WileyNJDv5} 
\usepackage{graphicx}
\usepackage{hyperref}
\usepackage{multirow}

\articletype{Research Article}%

\received{Date Month Year}
\revised{Date Month Year}
\accepted{Date Month Year}
\journal{Journal}
\volume{00}
\copyyear{2023}
\startpage{1}

\begin{document}

\title{Analytical description of the distributions of primary and secondary
  cosmogenic particles}

\author[1,2]{Shvetaank Tripathi}

\author[1,2]{Prashant Shukla}

\authormark{Tripathi \textsc{et al.}}
\titlemark{Analytical description of the distributions of primary and secondary
  cosmogenic particles}

\address[1]{\orgdiv{Nuclear Physics Division},
  \orgname{Bhabha Atomic Research Centre}, \orgaddress{Mumbai, \state{Maharashtra},
    \country{India}}}

\address[2]{\orgdiv{Homi Bhabha National Institute},
  \orgaddress{Mumbai, \state{Maharashtra},
    \country{India}}}

\corres{Corresponding authors: Prashant Shukla; Shvetaank Tripathi.
  \email{pshukla@barc.gov.in; shvetaank05@gmail.com}}



\abstract[Abstract]{
  In this work, we present an analytical description of the energy distributions of
  primary and secondary cosmogenic particles on Earth in terms of parameters having
  {clear} physical meaning.
  A modified power law is assumed for energy distributions, incorporating terms such as
  energy loss/decay, which are effective at low energies, and a source term, which is
  dominant at high energies.
  The parametrizations of the momentum distribution of primary protons and helium have been
  obtained including energy loss term.
  {For muons, both the energy loss and decay terms have been included.
    It is shown analytically that zenith angle distributions is given by $\cos^{n-1}\theta$
    in terms of energy index $n$ and the presence of decay term does not affect it.
    The analytical function describes the muon momentum distribution data at different
    altitudes and zenith angles.}
  The same form is also applied to describe the atmospheric muon and electron-type
  neutrino distributions simulated at various sites.
  The presented analytical functions provide an excellent description of all kinds of
  cosmogenic particles.
}


\jnlcitation{\cname{%
\author{Tripathi S}, and
\author{Shukla P}}.
}

\maketitle

\renewcommand\thefootnote{}
\footnotetext{\textbf{Abbreviations:} EAS, Extensive Air Shower.}

\section{Introduction} \label{intro}

Soon after their discovery~\cite{Hess:1912srp,1912NCim....3...93P,1570291225348853632,PhysRev.27.353,PhysRev.27.645,
  PhysRev.32.533,rossi1964cosmic}, the study of cosmic rays and their interactions with Earth's atmosphere
became an active area of research~\cite{10.1093/ptep/ptac097}. 
Numerous experiments have been conducted worldwide to investigate their origin, acceleration mechanisms, propagation,
and ionization {effects}~\cite{MAENO2001121,SHIKAZE2007154,2009BRASP..73..564P,2006astro.ph.12377P,2011Sci...332...69A,AGUILAR2002331,PhysRevLett.114.171103,PhysRevLett.115.211101,2009ApJ...707..593A,2010ApJ...714L..89A,Ryan:1972ek,BOCCIOLINI1996403,1997ApJ...487..415B,1999ApJ...518..457B,1998ApJ...502..278A}.
It is now understood that cosmic rays originate from astrophysical events such as supernovae, active galactic nuclei,
pulsars, and gamma-ray bursts, gaining energy via acceleration in intergalactic magnetic fields.
They strike Earth's atmosphere at a rate of approximately 1000 particles per square meter per second.
Most of the detected cosmic rays originate within our galaxy, as extra-galactic particles are largely deflected
by the Galactic magnetic field.
Additionally, the Sun’s magnetic field prevents low-energy particles with $E < 1~\mathrm{GeV}$ from reaching Earth.
Cosmic rays are ionized atomic nuclei, composed primarily of protons ($\sim$90\%) and alpha particles ($\sim$9\%),
with the remaining fraction consisting of heavier nuclei.

Upon entering Earth's atmosphere, cosmic ray particles undergo high-energy collisions with air nuclei.  
These interactions generate showers of secondary particles, predominantly pions and kaons, along with hyperons,
nucleon-antinucleon pairs, and small fractions containing charm quarks~\cite{grieder2001cosmic}.  
Subsequent decay of these secondary particles leads to the formation of Extensive Air Showers (EAS),
comprising hadronic, mesonic, and electromagnetic components.  
The shower contains both charged and neutral particles, many of these having finite lifetime but {still}
reach sea level due to relativistic effects.  
As the charged particles propagate, they continuously lose energy via processes like ionization and interactions
with nuclei.  The ionization losses decrease as the energy of the particle increases and become almost constant.
At very high energies the radiative losses become dominant.   
The electromagnetic cascades which result from pair production and bremsstrahlung, were formally described by
Bethe and Heitler~\cite{Bethe:1934za}.

The flux distribution for primary cosmogenic particles is observed to follow the power law, i.e.
\begin{equation}\label{eq:powlaw}
  \frac{dN}{dE} \propto E^{-n},
\end{equation}
where $n$ denotes the spectral index which carries the value $\approx$ 2.7 in wide energy range.
The power law is also applied to the secondary particles produced in shower.  
Use of such power-law spectrum may have helped in calculations regarding pion spectrum which led
to the formulation of the linear development of a cascade of particles in the atmosphere.
This work is extensively explained in~\cite{Gaisser:2016uoy}.

The power law grossly explains the momentum distribution of particles at high energies,
but at lower energies the effect of interactions must be taken into account.
Numerical calculations are normally performed to account accurately for decay and energy
loss processes along with the knowledge of primary cosmic spectrum and the
energy dependencies of their interaction cross-sections.
In recent years, several semi-analytical models~\cite{Gaisser:2016uoy,Gaisser:2001sd,Pan:2023gji} have
been proposed aiming to describe cosmogenic particles and their interactions in Earth's atmosphere.
Theoretical calculations~\cite{Allkofer:1971qr,Ayre:1973df,Lipari:1993hd} and
transport models~\cite{Maeda:1973nz,Stephens:1979ic} provided detailed investigations
regarding spectra for secondary cosmogenic particles. 
In addition, Monte Carlo codes such as CORSIKA~\cite{Heck:1998vt} employ different models
to calculate the interactions of cosmic particles in the Earth’s atmosphere~\cite{Sogarwal:2022hal}.
Recent advances in muon spectrum calculations have significantly enhanced our understanding
of cosmic ray interactions for underground~\cite{Fedynitch:2021ima,Azarkin:2024ned,Woodley:2024eln}
and atmospheric physics~\cite{sato2022measurement,PhysRevD.81.012001,adamson_2015_7sqzm-c3162}.
The IceCube Collaboration recently conducted an extensive analysis of seasonal variations
in the atmospheric muon neutrino spectrum, using 11.3 years of data, reporting
energy-dependent variations in agreement with theoretical predictions~\cite{2023EPJC...83..777A,2025EPJC...85.1368I}.

The aim of this work is to derive simple analytical formulae to describe the momentum
distributions of primary and secondary particles. These functions incorporate realistic
features of the distributions through parameters with clear physical significance.
The resulting expressions can be both practical and useful in various applications,
such as serving as input for detector simulations in cosmic ray experiments and searches
for rare physical phenomena.
Moreover, analytical expressions can be used to extrapolate measured data in unknown
regions and obtain integrated flux.
This work can be considered an extension of the study
presented in~\cite{Shukla:2016nio}.

For primary particles, we introduce an energy loss term that
is significant at lower energies but becomes negligible at higher energies.  
For secondary particles such as muons, we incorporate an exponentially decaying function to
account for their reduced abundance at lower energies. Additionally, we include dependencies
on key variables such as atmospheric depth and zenith angle.  
Using these formulations, we determine the values of the parameters with experimental data
collected worldwide for protons and helium~\cite{MAENO2001121,SHIKAZE2007154,2009BRASP..73..564P,2006astro.ph.12377P,2011Sci...332...69A,AGUILAR2002331,PhysRevLett.114.171103,PhysRevLett.115.211101,2009ApJ...707..593A,2010ApJ...714L..89A}, as well as for
muons~\cite{HAINO200435, Rastin:1984nu, MGardener_1962, PJHayman_1962, https://doi.org/10.1029/92JA02672, PhysRevD.19.1368, PhysRevLett.83.4241, Sogarwal:2022kgw}.  
Furthermore, we analyze variations in muon flux with atmospheric depth and zenith
angle using our modified formulation, fitting it to experimental data from
Bellotti~\cite{PhysRevD.53.35} et al. and the L3 collaboration~\cite{ACHARD200415}.  
We extend this formula to neutrinos by incorporating a source
term instead of a decay function, reflecting their stability and minimal interactions.  

The structure of this paper is as follows: Section~\ref{sec:ParamPrim} discusses the
description of the flux distribution for primary particles and comparison with previous results.  
Section~\ref{sec:ParamMuon} presents the description of energy and zenith angle distributions
for muons. Section~\ref{sec:VarMuon} presents studies of variations in muon flux with atmospheric
depth and zenith angle.  
In Section~\ref{sec:ParamNeut}, we extend the formulation to atmospheric neutrinos.  
Finally, Section~\ref{sec:Summ} provides a summary of our findings.

\section{Flux Distribution of Primary Cosmic Particles} \label{sec:ParamPrim}

The flux for primary and secondary cosmogenic particles as a function of energy has
been described in~\cite{Shukla:2016nio} as:
\begin{equation} \label{eq:Epara1}
    I(E) = I_0 N (E_0d + E)^{-n} \left(1 + \frac{E}{\epsilon}\right)^{-1},
\end{equation}
where $E_0d$ could be the energy loss by particles, $N$ is the normalisation constant obtained
numerically, $n$ represents the spectral index and $I_0$ is the integrated flux, given by
$I_0 = \int_{E_c}^{\infty} I(E) \, dE,$
with $E_c$ being the lower cut-off energy of the data.
The parameter $\epsilon$ was added to modify the power in the high energy part which
should account for the finite lifetimes of pions and kaons.

The parameters of Eq.~\ref{eq:Epara1}, as obtained by fitting the measured distributions
for primary protons and helium, have already been given in~\cite{Shukla:2016nio}.
This distribution does not give good distribution of data at lower energies where the
flux is suppressed. Hence, a new modified distribution function for primary particles is proposed:
\begin{equation} \label{eq:prohelflux}
    I(E) = I_0 N \left(\frac{E_1}{E} + E \right)^{-n}.
\end{equation}
Primary particles, such as protons and helium, might experience energy loss while
propagating through the cosmos and the upper layers of the atmosphere.
{The introduction of the term $E_1/E$ for primary particles is
  motivated by the energy loss experienced by massive charged particles as they
  traverse the atmosphere.
  At lower energies, the rate of energy loss through ionization and other electromagnetic
  processes increases approximately as $1/E$ as we move down in energy, leading to a
  suppression of the spectrum.
  The inclusion of the $E_1/E$ term effectively accounts for this low-energy attenuation. Here the unit of $E_1$ is GeV$^2$ .}

\begin{figure}[hbt!]
  \centering
  \includegraphics[width=8cm]{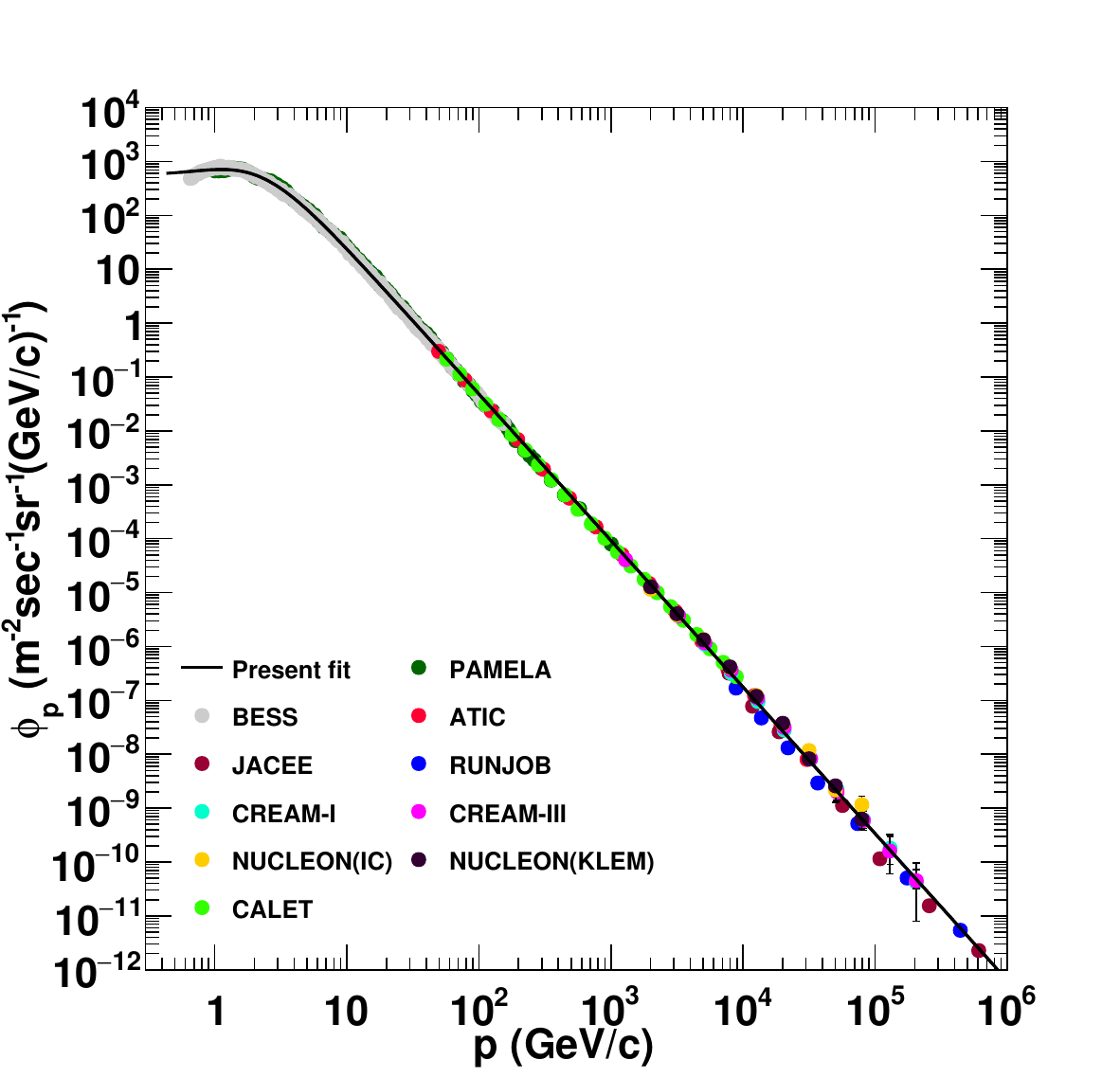}
  \includegraphics[width=8cm]{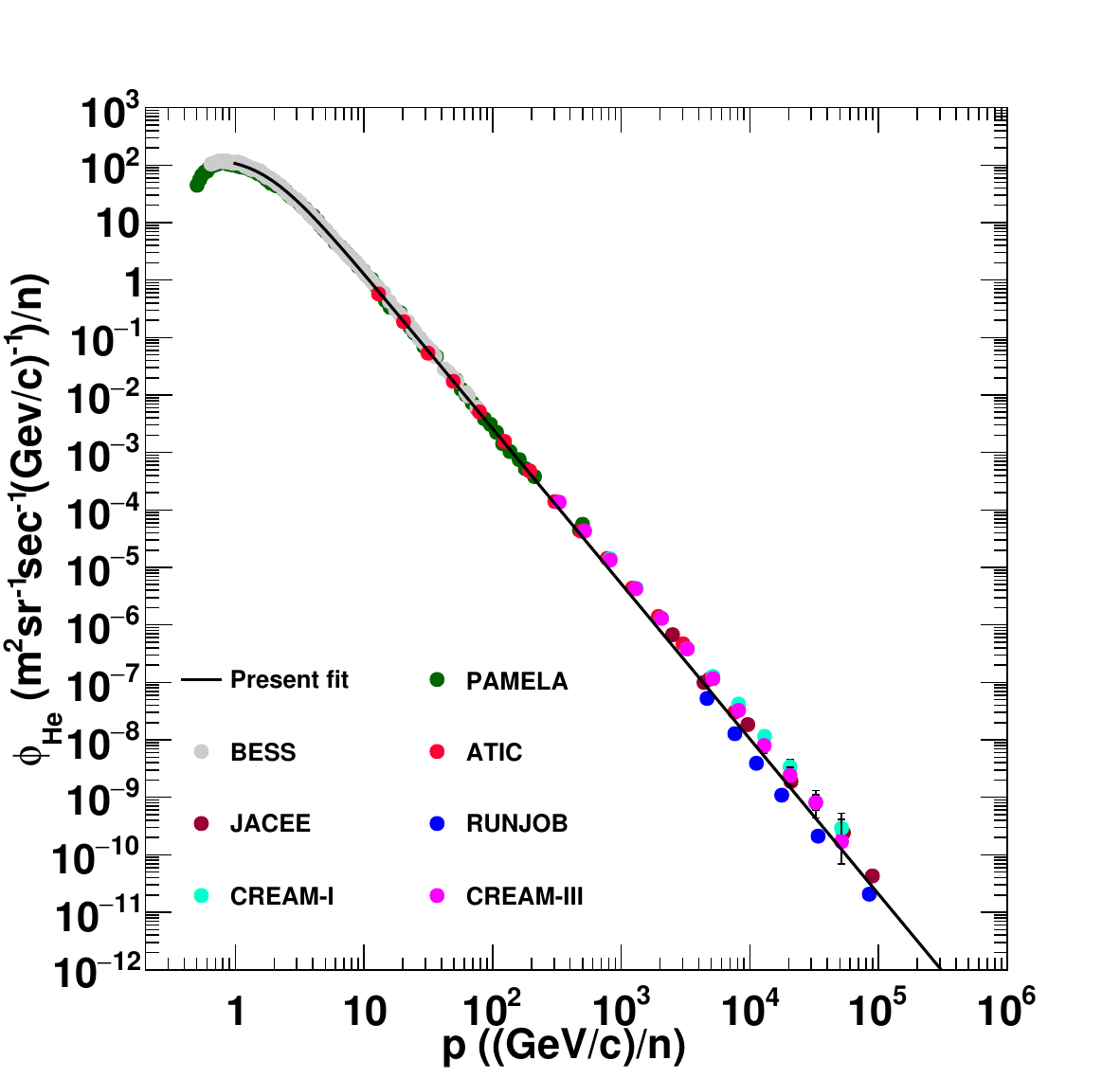}
  \caption{Proton and Helium flux~\cite{MAENO2001121,SHIKAZE2007154,2009BRASP..73..564P,2006astro.ph.12377P,2011Sci...332...69A,AGUILAR2002331,PhysRevLett.114.171103,PhysRevLett.115.211101,2009ApJ...707..593A,2010ApJ...714L..89A}
 as a function of momentum, fitted with Eq.~\ref{eq:prohelflux}.}
  \label{fig:prohelflux}
\end{figure}

Figure~\ref{fig:prohelflux} shows 
Proton and Helium flux~\cite{MAENO2001121,SHIKAZE2007154,2009BRASP..73..564P,2006astro.ph.12377P,2011Sci...332...69A,AGUILAR2002331,PhysRevLett.114.171103,PhysRevLett.115.211101,2009ApJ...707..593A,2010ApJ...714L..89A}
as a function of momentum, fitted with Eq.~\ref{eq:prohelflux}.
The parameters obtained by fitting proton and helium data are given in Table~\ref{tab:prohelflux}.
The value of spectral index, $n$ comes out to be near 2.7 for both the primary particles. 
The higher values of $\chi^2$/NDF can also be attributed to the fact that we have
combined data from multiple experiments, as no single experiment covers the entire energy range.
{This function provides a good description of energy distribution for primary
  particles with energy as low as 1 GeV.
  The exact production processes of primary particles and their interactions in the intergalactic
  media are difficult to be determined, especially the spectral shape at low energy.}

\begin{table*}[hbt!]
  \centering
  \caption{Parameters obtained by fitting proton and helium data with Eq.~\ref{eq:prohelflux}.}
  \begin{tabular*}{\textwidth}{@{\extracolsep\fill}llllll@{\extracolsep\fill}}
    \toprule
    & $I_{0}$ & $E_1$ & $n$ & $\chi^{2}/\rm{NDF}$ & Data Reference \\
    & (m$^{-2}$ s$^{-1}$ sr$^{-1})$ & (GeV)$^2$ &  &  &  \\
    \midrule
    Protons & 2606.68 {$\pm$ 36.87} & 2.16 $\pm$ 0.02 & 2.72 $\pm$ 0.04 & {10.02} & 
    ~\cite{MAENO2001121,SHIKAZE2007154,2009BRASP..73..564P,2006astro.ph.12377P,2011Sci...332...69A,AGUILAR2002331,PhysRevLett.114.171103,PhysRevLett.115.211101,2009ApJ...707..593A,2010ApJ...714L..89A}
     \\
    Helium & 289.24 {$\pm$ 4.96} & 0.80 $\pm$ 0.02 & 2.70 $\pm$ 0.04 & 5.81 &
    ~\cite{MAENO2001121,SHIKAZE2007154,2009BRASP..73..564P,2006astro.ph.12377P,2011Sci...332...69A,AGUILAR2002331,PhysRevLett.114.171103,PhysRevLett.115.211101,2009ApJ...707..593A,2010ApJ...714L..89A}
     \\
    \bottomrule
  \end{tabular*}
  \label{tab:prohelflux}
\end{table*}

\section{Flux Distribution of Atmospheric Muons} \label{sec:ParamMuon}

In this section, we study energy and zenith angle distribution of muons.
The distribution of secondary cosmogenic particles such as muons has been described by
Eq. \ref{eq:Epara1} which is given in~\cite{Shukla:2016nio}.
The parameters of the distribution, corresponding to data collected from various
experiments are also given in~\cite{Shukla:2016nio}.
{The comparison of measured data for muons at various locations
  along with parametrized function (Eq.~\ref{eq:Epara1}) with various combinations of
  low-energy and high-energy models (GEISHA \& QGSJET, UrQMD \& QGSJET, UrQMD \&
  QGSJET-II, GEISHA \& SIBYLL, UrQMD \& SIBYLL and FLUKA \& SIBYLL)
  was done in Ref.~\cite{Sogarwal:2022hal}(see fig. 2, 3, 4).}
This work attempted to use a modified power law which works at lower energies. 
However, it does not account for the decaying nature of particles like muons.  
While high-energy muons can travel large distances, lower-energy muons tend to decay
into electrons or positrons, producing the neutrinos and anti-neutrinos.

\subsection{Energy Distribution} \label{subsec:Erg}

We propose a modified energy distribution function to describe muon momentum distribution 
which includes effect of decay of muons:
\begin{equation} \label{eq:muonflux}
    I(E) = I_0 N \left(\frac{E_0 d}{\cos \theta} + E\right)^{-n} \exp \left(-\frac{k}{E \cos \theta}\right), 
\end{equation}
where variables $d$ and $\theta$ represent the atmospheric depth and zenith angle, respectively,
ensuring that the function captures variations in flux distribution with respect to these variables.
The exponential term accounts for decay of muons which can be understood as follows.
A particle travelling with velocity $\beta$ will cover a distance $\beta \gamma \tau c$ during
its lifetime $\tau$. 
This leads to the decay rate per mean free path, $\lambda_d$, expressed as  
\begin{equation}
    \frac{1}{\lambda_d} = \frac{1}{\beta \gamma \tau c} = \frac{m}{\tau p c},
\end{equation}
where $m$ is the mass of the particle and $p$ is its momentum.  

Using this, we compute the survival probability of the particle after traversing a thickness $X$ along an
inclined trajectory with zenith angle $\theta$:  
\begin{equation}\label{eq:Surv}
    S = \exp \left(-\int \frac{m}{\tau p c \cos \theta} dX\right).
\end{equation}
For muons, the survival probability at a given energy, $E$ in a fixed small distance, $dX$ can be written as:   
\begin{equation}\label{eq:Surv2}
    S \approx \exp \left(-\frac{k}{E \cos \theta}\right),
\end{equation}
where $k$ is a parameter encapsulating all constant factors ($m, \tau$) leaving only $E$ and $\cos\theta$.  
The parameter $k$ depends on the mass and lifetime of the particle 
and governs the flux loss at lower energies.
{
  The exponential term in the proposed analytical function reflects the same physical mechanism
  predicted by cascade theory, the progressive depletion of low energy muons due to decay before
  reaching the observation level, while at higher energies, the contribution becomes negligible,
  leaving the power-law behaviour dominant.
  For vertical muons, the exponential term becomes $\sim$1 above energy 18 - 20 GeV.
  It also depends upon the zenith angles, eg. at 60$^{\circ}$ zenith angle range, the energy at which the
  exponential terms become $\sim$1 increases to $\sim$ 35 - 40 GeV.}

\begin{figure}[hbt!]
  \centering
  \includegraphics[width=7.5cm]{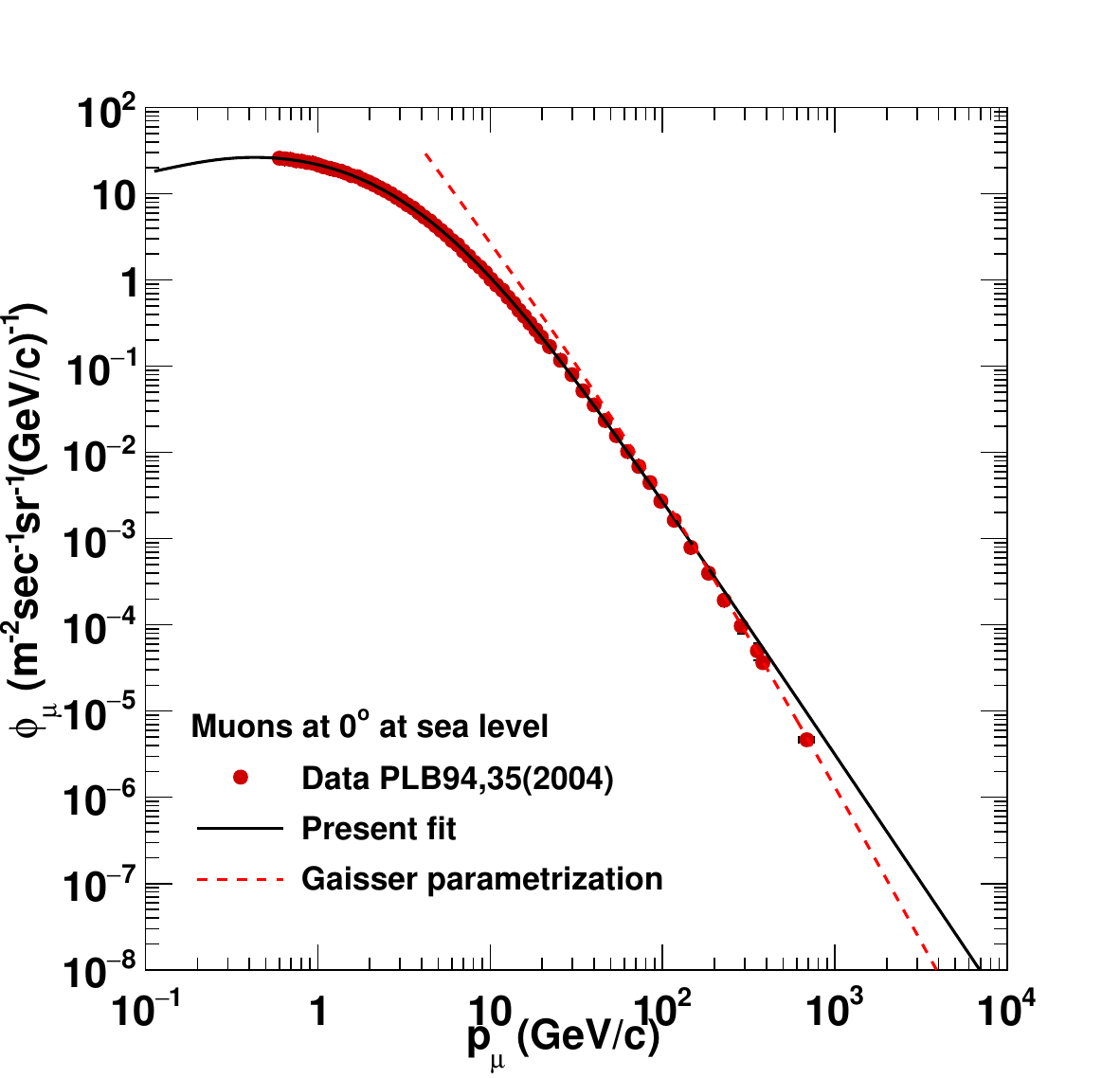}
  \includegraphics[width=7.5cm]{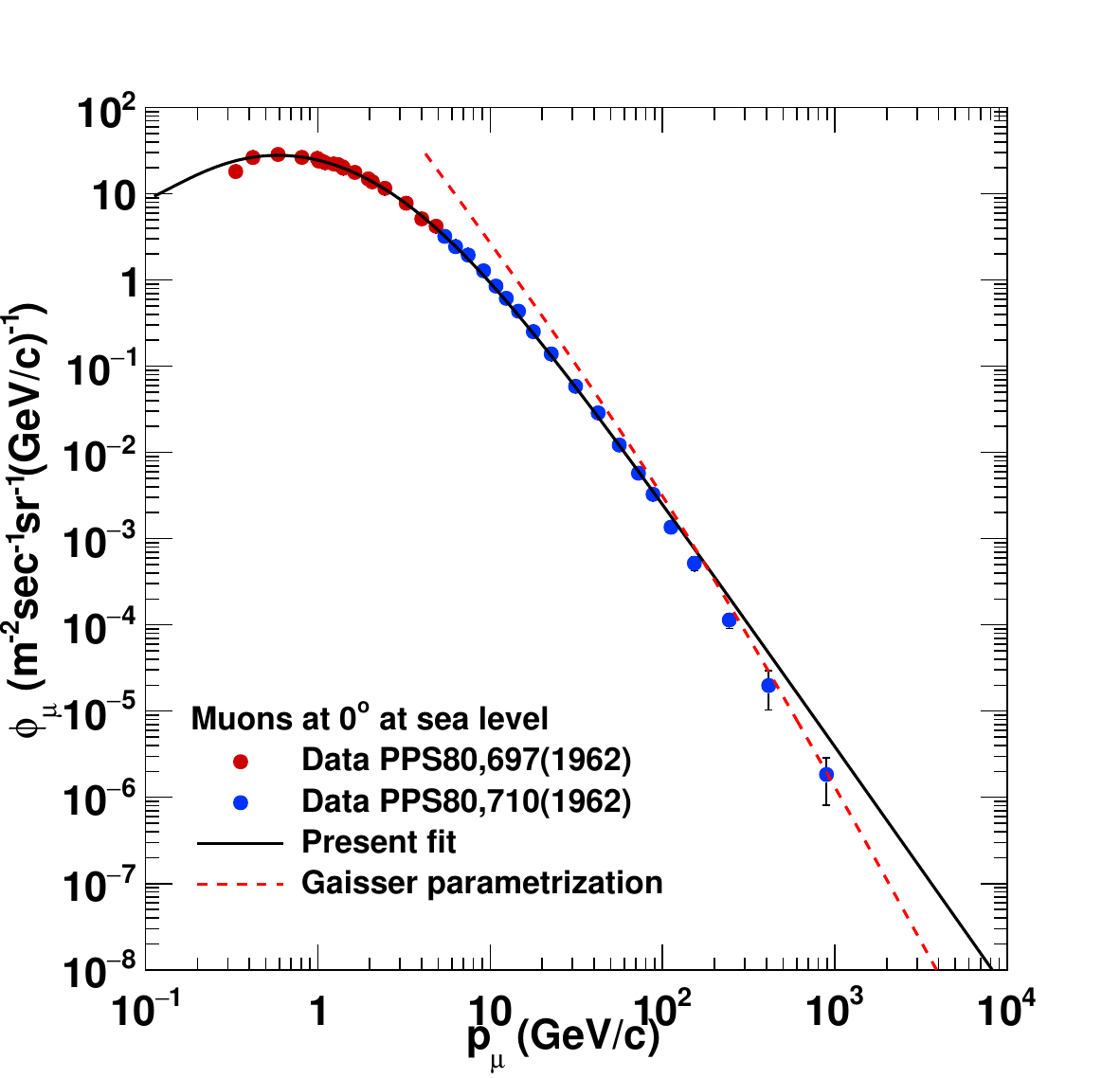}
  \includegraphics[width=7.5cm]{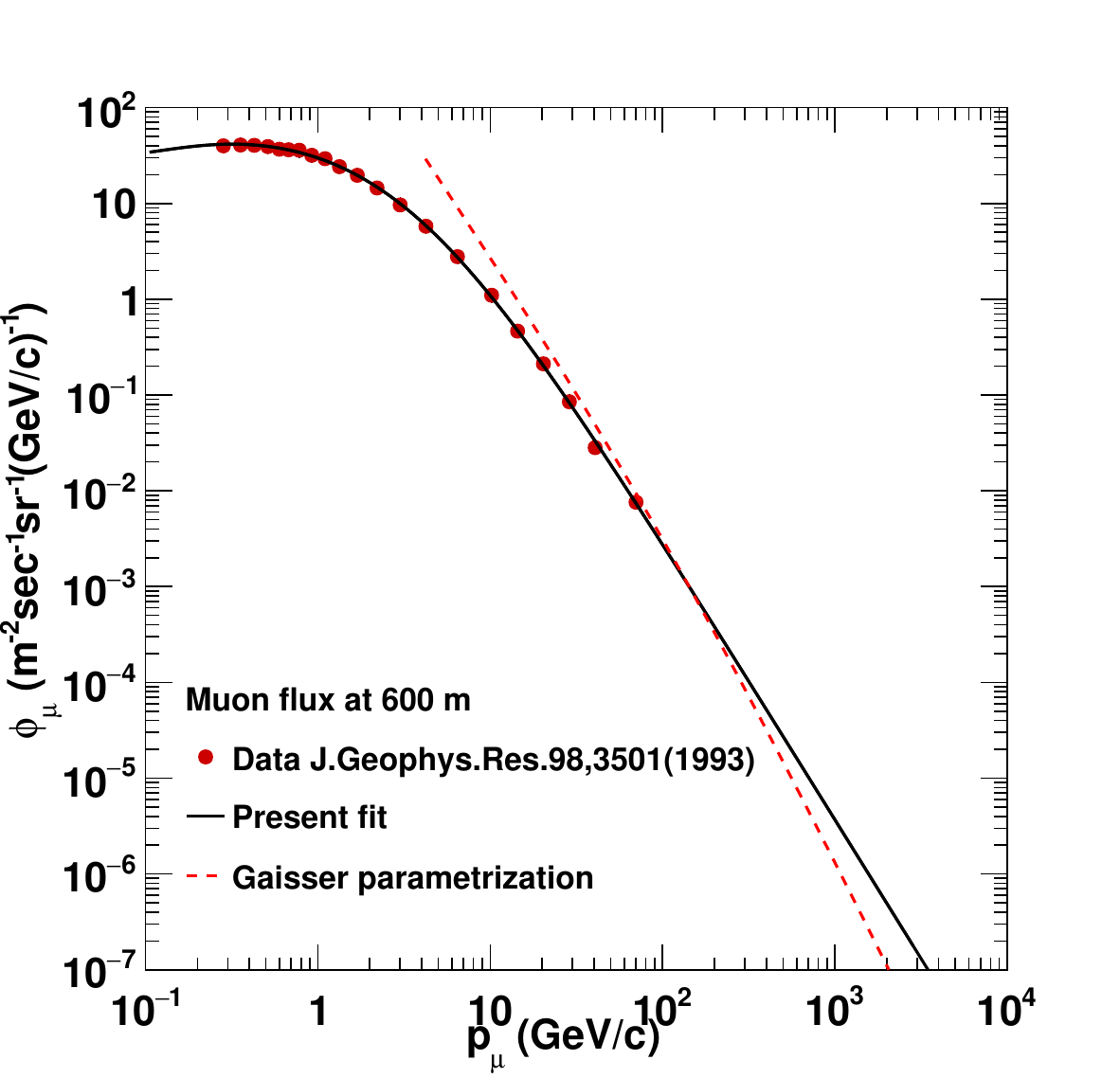}
  \includegraphics[width=7.5cm]{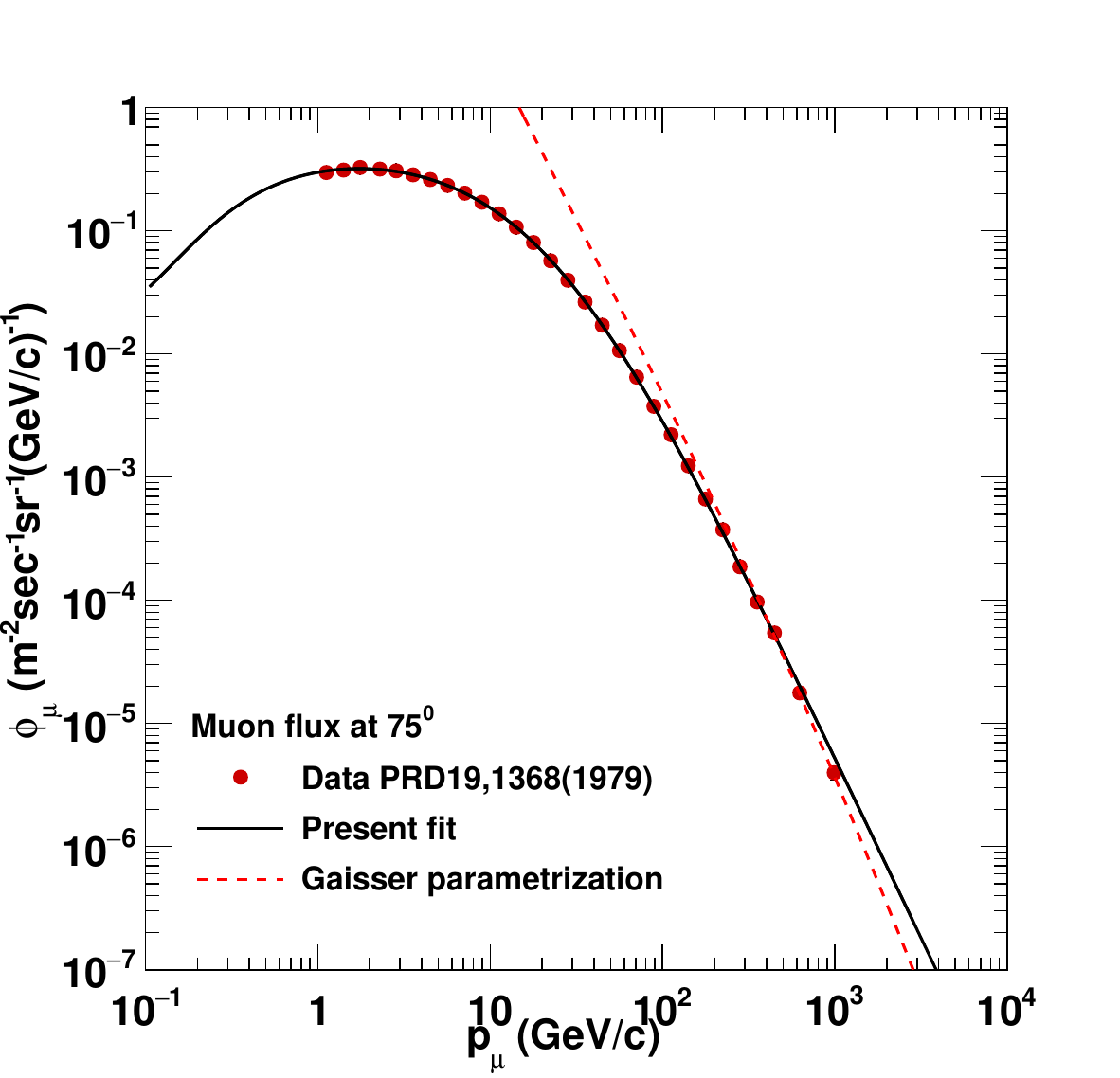}
  \includegraphics[width=7.5cm]{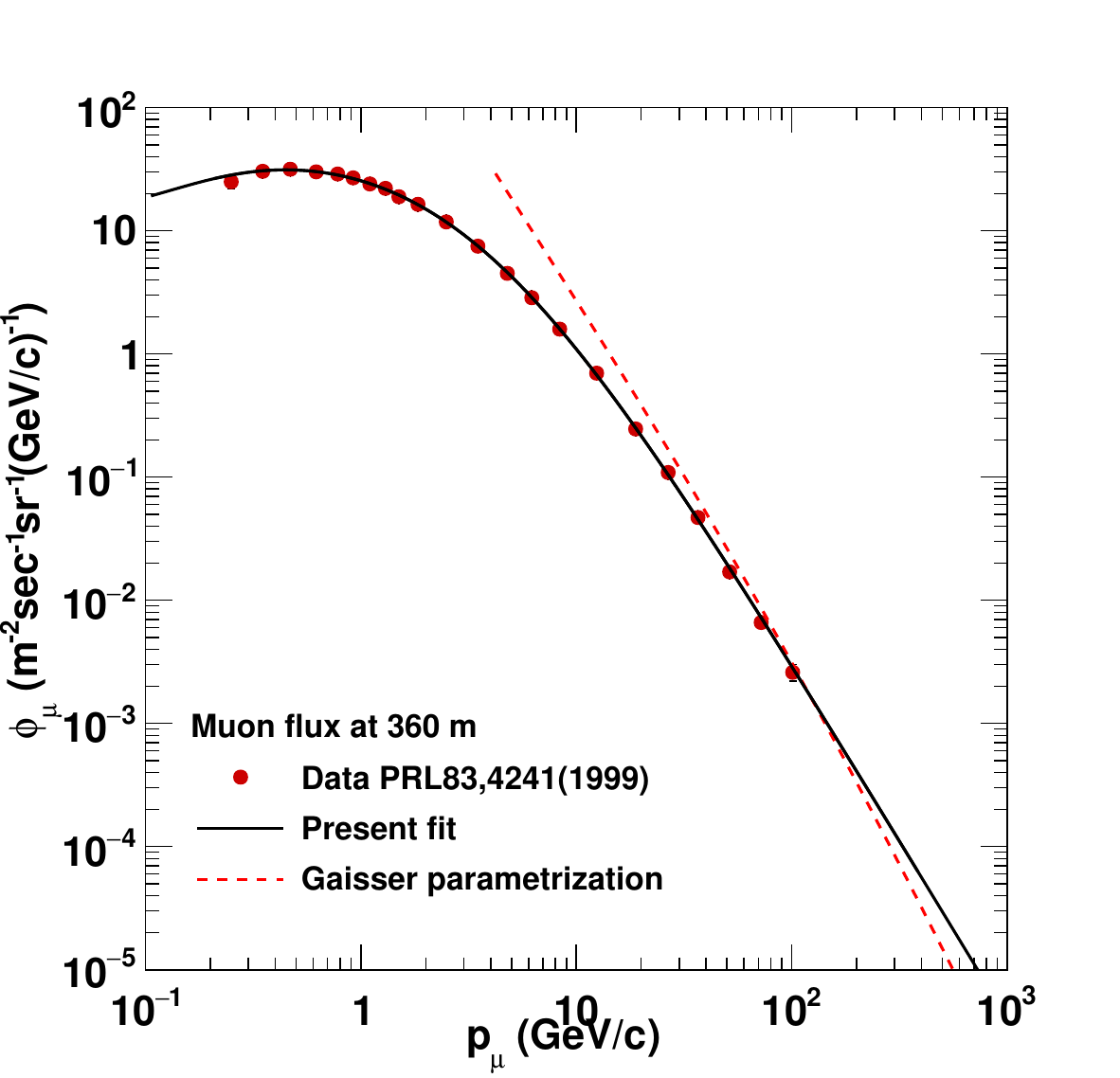}
  \includegraphics[width=7.5cm]{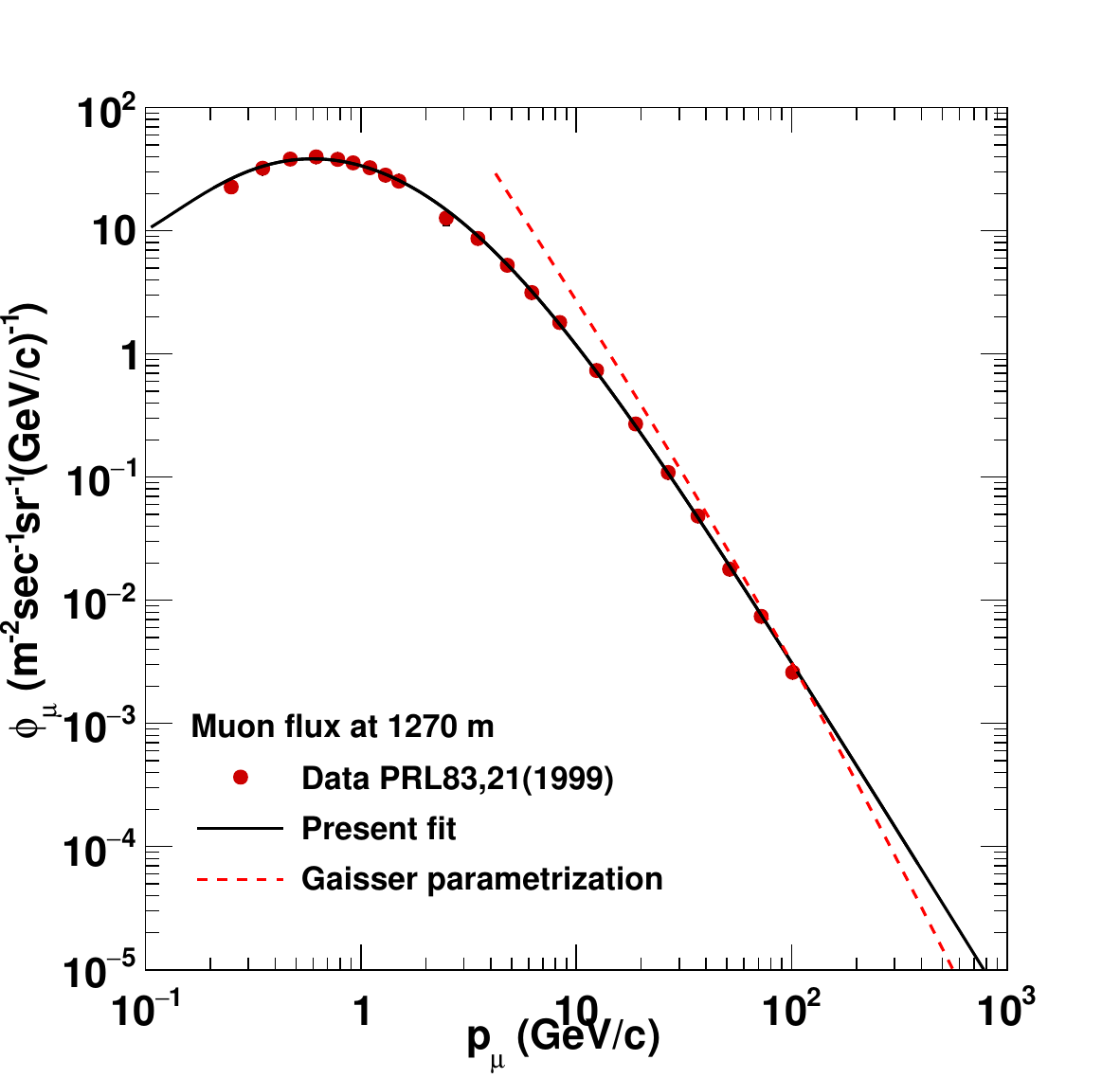}
  \caption{Muon flux as a function of momentum at various locations~\cite{HAINO200435, Rastin:1984nu, MGardener_1962, PJHayman_1962, https://doi.org/10.1029/92JA02672, PhysRevD.19.1368, PhysRevLett.83.4241, Sogarwal:2022kgw}, fitted with Eq.~\ref{eq:muonflux}.}
  \label{fig:muonflux}
\end{figure}

Figure~\ref{fig:muonflux} shows the muon flux as a function of momentum at various
locations~\cite{HAINO200435, Rastin:1984nu, MGardener_1962, PJHayman_1962, https://doi.org/10.1029/92JA02672, PhysRevD.19.1368, PhysRevLett.83.4241, Sogarwal:2022kgw}, fitted with Eq.~\ref{eq:muonflux}. 
We have also included the parametrization done by T.K. Gaisser
(see Ch.6 of~\cite{Gaisser:2016uoy}), which gives a good fitting at higher momentum
but fails at lower momentum. In contrast, the present function provides a better fit in the
whole range of momentum.
The values of parameters obtained after fitting muon distributions with Eq.~\ref{eq:muonflux},
keeping $E_0.d$ and $k$ as free, are given in Table~\ref{tab:muonflux}.
{Comparison between present fit (Eq.~\ref{eq:muonflux}) and Gaisser Parametrization
  is also shown in terms of $\chi^{2}/\rm{NDF}$ values for both models.}
{The spectral index $n$ is slightly larger than that observed for primary particles.}
The value of $k$ lies between 0.10 to 0.35 GeV.
It can have a correlation with $E_0d$. The value of $E_0d$ in general
decreases with increasing altitude and increases with zenith angle.
This analysis helps to obtain the integrated flux, given by $I_0$, from measured
data which requires extrapolation.
{The function given by Eq.~\ref{eq:muonflux} works well in case of muons in
  the low energy range, as low as 200 MeV.
  Data for muons below this energy are not available which can be due to measurement difficulties.}

\begin{table*}[hbt!]
  \centering
  \caption{Parameters obtained after fitting muon distributions with Eq.~\ref{eq:muonflux}, keeping $E_0.d$ and $k$ as free.
    {Comparison between present fit (Eq.~\ref{eq:muonflux}) and Gaisser Parametrization~\cite{Gaisser:2016uoy} is shown in terms
  of $\chi^{2}/\rm{NDF}$ values for both models.}}
  \begin{tabular*}{\textwidth}{@{\extracolsep\fill}llllllll@{\extracolsep\fill}}
      \toprule
      & $I_{0}$     & $E_{0}.d$  & $n$      &   $k$   &\multicolumn{2}{c}{$\chi^{2}/\rm{NDF}$} & Data \\         
      &   (m$^{-2}$ s$^{-1}$ sr$^{-1})$ & (GeV)  &  &  (GeV)  & Present Fit & {Gaisser Param.} & Reference\\
      \midrule
      $\mu$ at 0$^\circ$ & 81.35  & 3.82 & 2.98    & 0.14 & 0.90 & {8.90}   & Tsukuba, Japan\\
      at sea level  &{$\pm$ 4.28}  & $\pm$0.07 & $\pm$ 0.01     & $\pm$0.02 & & &(36.2$^{\circ}$ N, 140.1$^{\circ}$ W)~\cite{HAINO200435} \\
      $\mu$ at 0$^\circ$ & 80.23  & 2.58  & 2.84   & 0.32 & 12.56 & {63.34}    & Durham, UK\\
      at sea level  & {$\pm$ 3.76} & $\pm$0.05 & $\pm$ 0.01     & $\pm$0.01 & &  &(54.76$^{\circ}$ N, 1.57$^{\circ}$ W)~\cite{MGardener_1962, PJHayman_1962} \\
      $\mu$ at 0$^\circ$ & 103.69    & 3.06 &  2.90   & 0.10 &  1.91 &   {72.28}   & Prince Albert, Canada\\
      at $600$ m  & {$\pm$ 13.12}  & $\pm$ 0.13 & $\pm$ 0.04     & $\pm$0.02 & &  & (53.2$^{\circ}$ N, 105.75$^{\circ}$ W)~\cite{https://doi.org/10.1029/92JA02672}\\
      $\mu$ at 75$^\circ$ &  4.83  & 5.39 & 2.95   & 0.10 & 1.01 & {37.02}    & Hamburg, Germany\\
      at sea level  & {$\pm$ 0.44}  &   $\pm$ 0.09& $\pm$ 0.02     & $\pm$ 0.01 & &   & (53.56$^{\circ}$ N, 10$^{\circ}$ E)~\cite{PhysRevD.19.1368} \\
      $\mu$ at 0$^\circ$ &  90.10  & 3.24 & 2.90   & 0.16 & 0.728 & {71.54}     & Lynn Lake, Canada\\
      at 360 m & {$\pm$ 2.48}  &   $\pm$ 0.12& $\pm$ 0.03     & $\pm$0.02 & &    & (56.5$^{\circ}$ N, 101.0$^{\circ}$ W)~\cite{PhysRevLett.83.4241} \\
      $\mu$ at 0$^\circ$ &  106.69  & 2.36 & 2.82   & 0.35 & 1.981  & {8.89}     & Fort Summer, New Mexico\\
      at 1270 m & {$\pm$ 2.64}  &   $\pm$ 0.09 & $\pm$ 0.02     & $\pm$0.02 & &  & (34.3$^{\circ}$ N, 104.1$^{\circ}$ W)~\cite{PhysRevLett.83.4241}\\
      \bottomrule
  \end{tabular*}
  \label{tab:muonflux}
\end{table*}

\subsection{Zenith Angle Distribution} \label{subsec:Ang}
{
  The modified energy distribution function for muons, as defined in the previous subsection,
  also depends on the zenith angle, $\theta$.  
  Using Eq.~\ref{eq:muonflux}, we can obtain the energy-integrated flux, $\phi(\theta)$,
  at a given zenith angle $\theta$ as:
  \begin{equation}
    \phi(\theta) = I_0 N \int_{0}^{\infty}\left(\frac{E_0 d}{\cos \theta} + E\right)^{-n} \exp \left(-\frac{k}{E \cos \theta}\right) dE.
  \end{equation}
  Substituting, $E\cos\theta = x$ and $dE = dx/\cos\theta$, we have 
  \begin{equation}
    \phi(\theta) = \cos^{n-1}\theta \times I_0 N \int_{0}^{\infty}\left(E_0 d + x \right)^{-n} \exp \left(-\frac{k}{x}\right) dx =  \cos^{n-1} \theta \times \phi(0), 
  \end{equation}
  where $\phi(0)$ is the vertical intensity of muons. 
  This gives an analytical expression for the ratio of energy-integrated flux for muons coming at zenith angle,  $\theta$ with that for vertical muons as
  \begin{equation}
    \frac{\phi(\theta)}{\phi(0)} = \cos^{n-1}\theta.
  \end{equation}
  This shows that the decay term in the muon flux distribution does not affect
  the zenith angle distribution which effectively gets canceled out in the ratio.
  With energy index $n\sim 3$, we get the $\cos^2\theta$ distribution which is empirically
  shown to hold good with various measurements over time. }

\section{Variation in Muon Fluxes with respect to Atmospheric Depth and Zenith Angle} \label{sec:VarMuon}

The modified function for the distribution of muons given by Eq.~\ref{eq:muonflux}
includes dependencies on atmospheric depth ($d$) and zenith angle ($\theta$).
We use this function to describe negative muon flux as a function of momentum measured at
different atmospheric depths~\cite{PhysRevD.53.35} and obtain
the parameters $E_0$ and $k$. The value of $\theta$ is taken to be 0$^\circ$ while
studying the variation with atmospheric depth.
 The L3 experiment~\cite{ACHARD200415} measured distributions of both the positive and negative muons at
different zenith angles at an altitude 450 m above sea level which corresponds to
a value of $d=967$ g/cm$^{2}$. We use this data to obtain the parameters of Eq.~\ref{eq:muonflux}
as a function of zenith angle.

\subsection{Variation with respect to atmospheric depth :} \label{subsec:1}

\begin{figure}[hbt!]
  \centering
  \includegraphics[width=9cm]{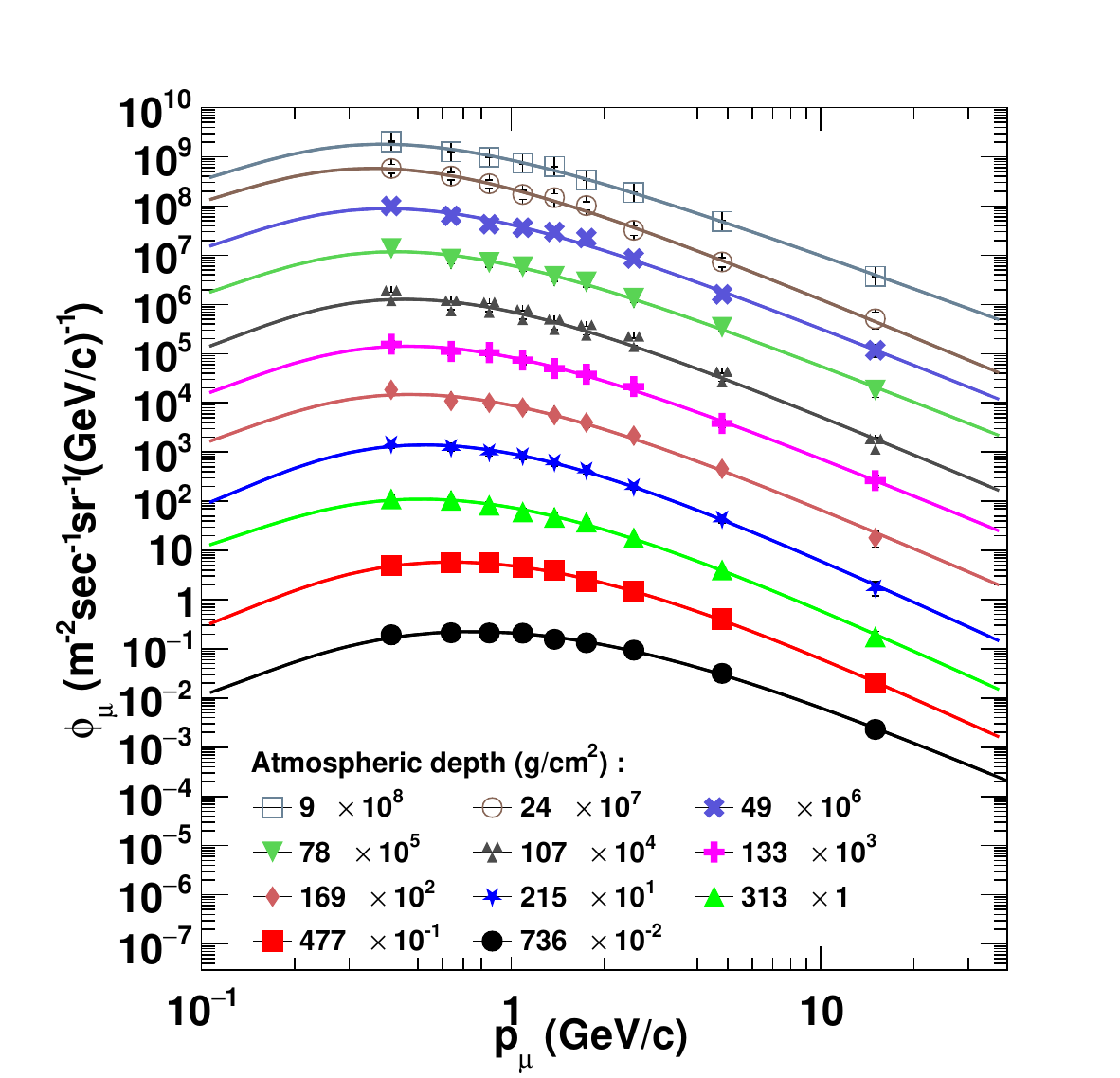}
  \caption{The negative muon flux as a function of momentum measured at different atmospheric depths~\cite{PhysRevD.53.35}.
   The solid lines are obtained by fitting the data with 
   Eq.~\ref{eq:muonflux} keeping $E_0$ = 0.002 GeV/g-cm$^{-2}$ and $k$ free. 
  } 
  \label{fig:FixedE0Dep}
\end{figure}

In this section, we analyze the muon flux distributions at various atmospheric depths.
First, the value of $E_0$ is kept fixed to be 0.002 GeV/g-cm$^{-2}$ which is close to the
value of energy loss for minimum ionizing particles and $k$ is kept free.
Figure~\ref{fig:FixedE0Dep} shows the negative muon flux as a function of momentum measured at
different atmospheric depths~\cite{PhysRevD.53.35}.
{In this analysis, the zenith angle, $\theta$ is fixed at 0$^{\circ}$.}
The solid lines are obtained by fitting the data with Eq.~\ref{eq:muonflux} keeping
$E_0$ = 0.002 GeV/g-cm$^{-2}$ and $k$ free. 
Table \ref{tab:FixedE0Dep} shows the values of parameters obtained.

\begin{table*}[hbt!]
  \centering
  \caption{The values of parameters obtained by fitting the data of 
    negative muon momentum distribution at different atmospheric depths~\cite{PhysRevD.53.35}
    with $E_{0}$ = 0.002 GeV/g-cm$^{-2}$ and $k$ free.}
  \begin{tabular*}{\textwidth}{@{\extracolsep\fill}llllll@{\extracolsep\fill}}
    \toprule
    Atmospheric depth & $I_{0}$       & $E_{0}$ & $n$     & $k$     & $\chi^{2}$/\rm{NDF} \\ 
    (g/cm$^{2}$)    &  (m$^{-2}$s$^{-1}$sr$^{-1}$) & (GeV/g-cm$^{-2}$)  &  & (GeV) & \\
    \midrule
    736 & 60.521 {$\pm$ 24.585} & 0.002 & 2.70 $\pm$0.19 & 0.695 $\pm$0.236 & 0.122 \\
    477 & 112.119 {$\pm$ 32.802} & 0.002 & 2.94 $\pm$0.12 & 0.779 $\pm$0.166 & 0.177 \\
    313 & 164.766 {$\pm$ 44.553} & 0.002 & 2.91 $\pm$0.15 & 0.691 $\pm$0.164 & 0.248 \\
    215 & 196.095 {$\pm$ 53.801} & 0.002 & 2.93 $\pm$0.16 & 0.872 $\pm$0.182 & 0.258 \\
    169 & 198.829 {$\pm$ 57.901} & 0.002 & 2.75 $\pm$0.16 & 0.788 $\pm$0.204 & 0.890  \\
    133 & 191.673 {$\pm$ 53.564} & 0.002 & 2.62 $\pm$0.16 & 0.808 $\pm$0.197 & 0.485 \\
    107 & 161.134 {$\pm$ 45.232} & 0.002 & 2.68 $\pm$0.16 & 0.856 $\pm$0.206 & 0.813  \\
    78 & 148.316 {$\pm$ 40.189} & 0.002 & 2.51 $\pm$0.15 & 0.812 $\pm$0.200 & 0.527 \\
    49 & 101.676 {$\pm$ 26.034} & 0.002 & 2.55 $\pm$0.15 & 0.847 $\pm$0.191 & 1.263 \\
    24 & 56.986 {$\pm$ 18.329} & 0.002 & 2.64 $\pm$0.21 & 0.866 $\pm$0.253 & 0.502 \\
    9 & 22.121 {$\pm$ 0.847} & 0.002 & 2.30 $\pm$0.02 & 0.879 $\pm$0.032 & {0.168} \\
    \bottomrule
  \end{tabular*}
  \label{tab:FixedE0Dep}
\end{table*}

\begin{figure}[hbt!]
  \centering
  \includegraphics[width=9cm]{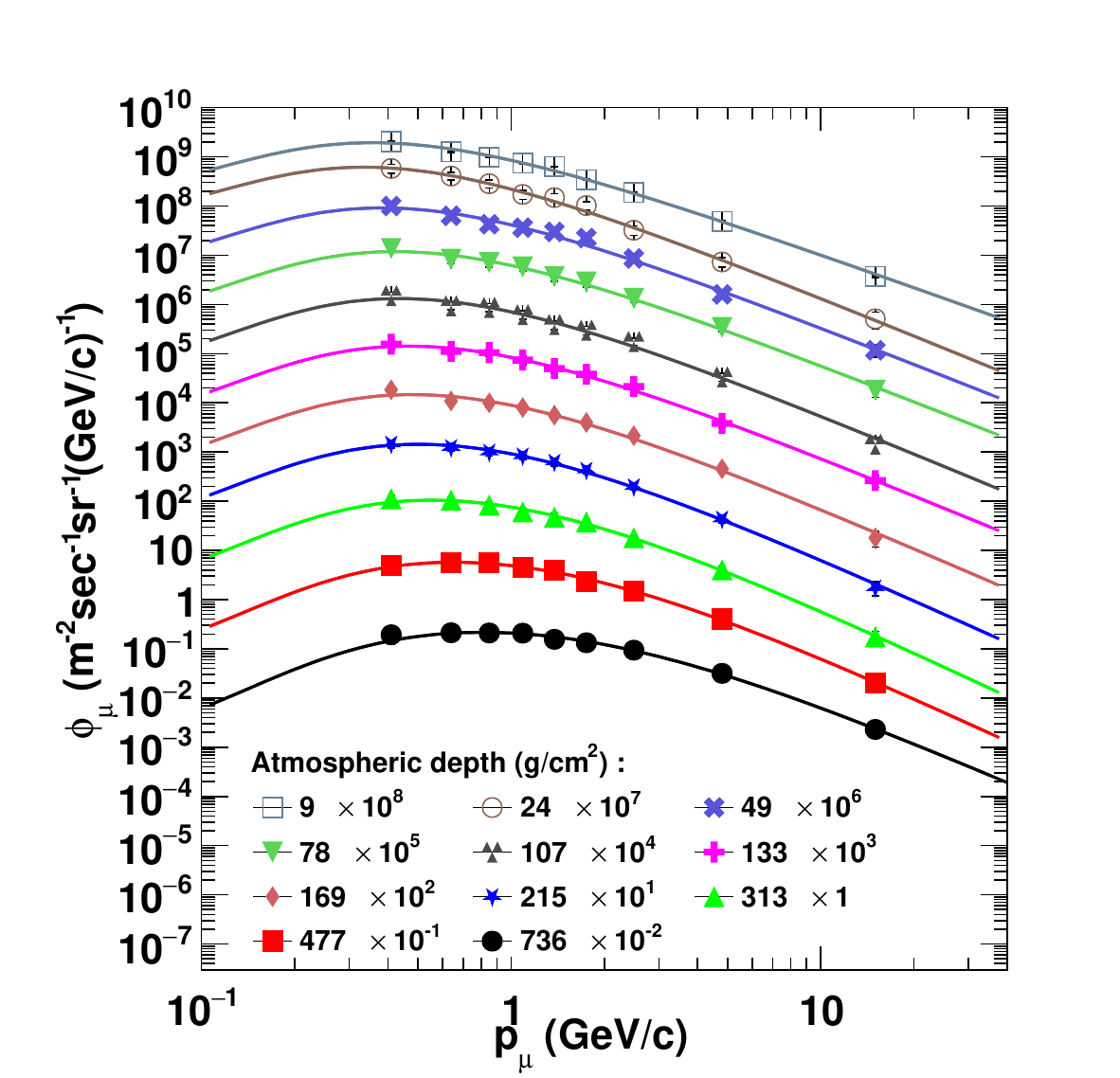}\\
  \caption{The negative muon flux as a function of momentum measured at different atmospheric depths \cite{PhysRevD.53.35}.
    The solid lines are obtained by fitting the data with Eq. \ref{eq:muonflux} keeping $E_0$ =
    0.002 GeV/g-cm$^{-2}$ and $k$ = 0.80 GeV. }
  \label{fig:FixedE0kDep}
\end{figure}

{The value of $k$, as substituted in Eq.~\ref{eq:Surv2}, depends on the total
  distance travelled by a muon, along with the constants such as mass and lifetime of muon and
  speed of light. This gives a value of $k$ between 0.16 to 1.6 GeV for muons travelling the
  distance of 1 to 10 kms.}
The value of $k$ obtained by fitting lies between 0.69 GeV to 0.88 GeV with small variation
and hence we fix it to a value 0.80 GeV.
Figure~\ref{fig:FixedE0kDep} shows the negative muon flux as a function of momentum at
different atmospheric depths~\cite{PhysRevD.53.35}.
The solid lines are obtained by fitting the data with Eq.~\ref{eq:muonflux} keeping
$E_0$ = 0.002 GeV/g-cm$^{-2}$ and $k$ = 0.80 GeV. 
Table~\ref{tab:FixedE0kDep} shows the values of parameters obtained.
The parametrization allows calculation of the integrated flux, $I_0$. 
In case of depth variation, it is the integrated vertical flux for the negative muon
and should be around half of the total muon integrated flux.
At a depth of 736 g/cm$^{2}$ (altitude $\sim$3 km), the flux is 59.89 m$^{-2}$s$^{-1}$sr$^{-1}$,
higher than that at sea level. 
The integrated flux $I_0$ first increases with increasing depth and then after reaching a certain
point, it begins to decrease. Additionally, the value of $n$ is 2.76 near the
Sea level, but it increases slightly with altitude before decreasing again, indicating that the muon
spectrum hardens at higher altitudes.

\begin{table*}[t!]
  \centering
  \caption{The values of parameters obtained by fitting the data of 
    negative muon momentum distribution at different atmospheric depths \cite{PhysRevD.53.35}
    with $E_{0}$ = 0.002 GeV/g-cm$^{-2}$ and $k$ = 0.80 GeV. }
  \begin{tabular*}{\textwidth}{@{\extracolsep\fill}llllll@{\extracolsep\fill}}
    \toprule
    Atmospheric depth & $I_{0}$       & $E_{0}$ & $n$     & $k$     & $\chi^{2}$/\rm{NDF} \\ 
    (g/cm$^{2}$)    &  (m$^{-2}$s$^{-1}$sr$^{-1}$) & (GeV/g-cm$^{-2}$)  &  & (GeV) & \\
    \midrule
    736 & 59.887 {$\pm$ 11.437} & 0.002 & 2.76 $\pm$0.12 & 0.80 & 0.131 \\
    477 & 111.848 {$\pm$ 13.434} & 0.002 & 2.96 $\pm$0.09 & 0.80 & 0.557 \\
    313 & 161.489 {$\pm$ 15.489} & 0.002 & 2.99 $\pm$0.09 & 0.80 & 0.273 \\
    215 & 198.513 {$\pm$ 15.905} & 0.002 & 2.88 $\pm$0.08 & 0.80 & 0.244 \\
    169 & 198.312 {$\pm$ 16.020} & 0.002 & 2.75 $\pm$0.08 & 0.80 & 0.763 \\
    133 & 191.993 {$\pm$ 15.076} & 0.002 & 2.62 $\pm$0.08 & 0.80 & 0.416 \\
    107 & 163.274 {$\pm$ 12.346} & 0.002 & 2.64 $\pm$0.07 & 0.80 & 0.707 \\
    78 & 148.761 {$\pm$ 11.190} & 0.002 & 2.50 $\pm$0.07 & 0.80 & 0.451 \\
    49 & 102.989 {$\pm$ 6.425} & 0.002 & 2.51 $\pm$0.06 & 0.80 & 1.091 \\
    24 & 58.465 {$\pm$ 4.329} & 0.002 & 2.59 $\pm$0.08 & 0.80 & 0.440 \\
    9 & 22.762 {$\pm$ 2.295} & 0.002 & 2.26 $\pm$0.01 & 0.80 & {0.153} \\
    \bottomrule
  \end{tabular*}
  \label{tab:FixedE0kDep}
\end{table*}

\subsection{Variation with respect to zenith angle :} \label{subsec:2}
\begin{figure}[t!]
  \centering
  \includegraphics[width=9cm]{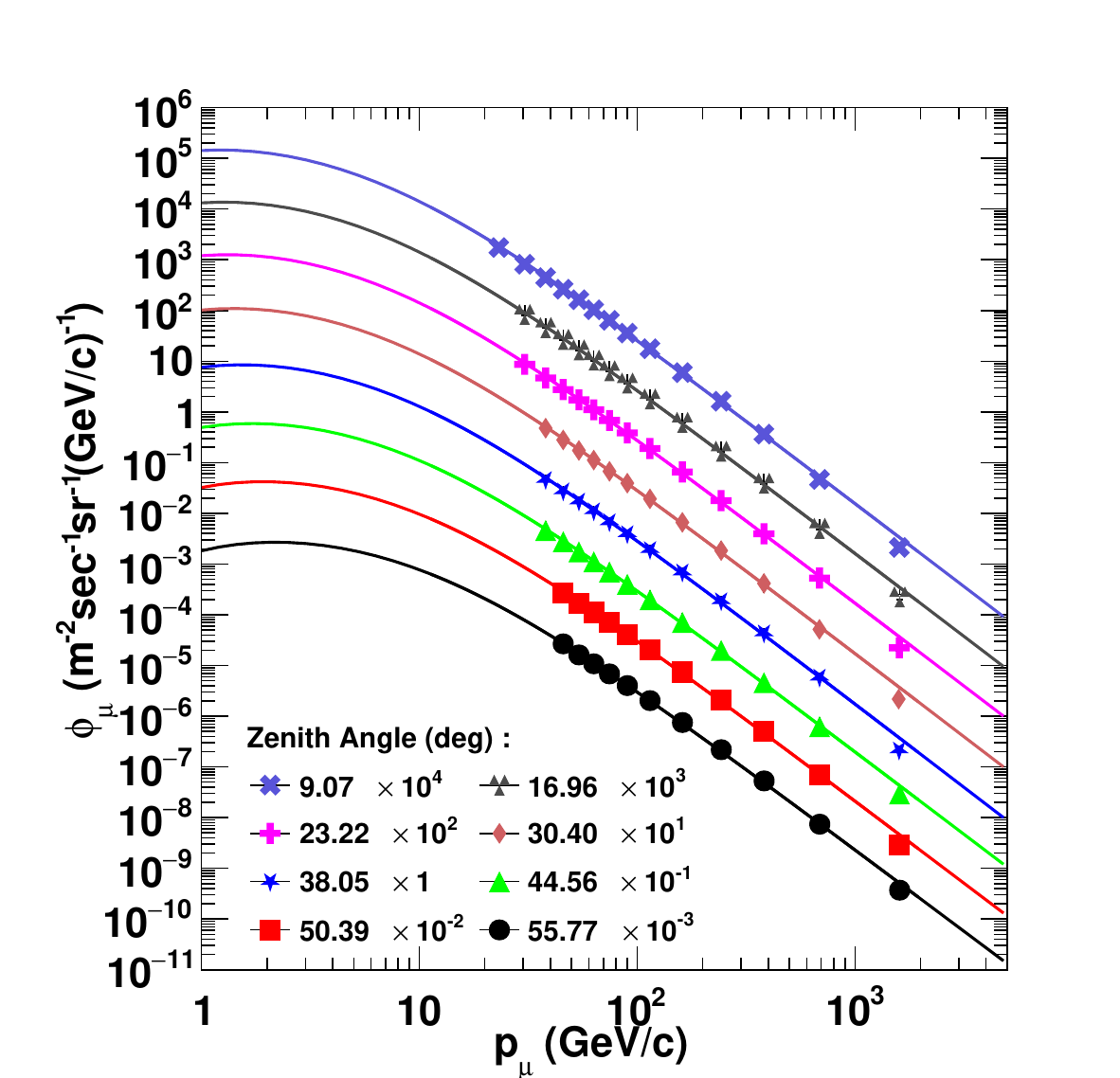}
  \caption{The total muon flux at different
    zenith angles~\cite{ACHARD200415} at an altitude of 450 m is given in right figure.
    The solid lines are obtained by fitting the data with Eq.~\ref{eq:muonflux}
    keeping $E_0$ = 0.005 GeV/g-cm$^{-2}$ and $k$ = 0.80 GeV.}
  \label{fig:FinalFixDepZen}
\end{figure}

\begin{table*}[hbt!]
  \centering
  \caption{The values of parameters obtained by fitting the total muon momentum distribution at
    different zenith angles~\cite{ACHARD200415} with $E_0$ = 0.005 GeV/g-cm$^{-2}$ and $k$ = 0.80 GeV. {Atmospheric depth, $d$ is fixed at 967.0 g/cm$^{2}$.}}
  \begin{tabular*}{\textwidth}{@{\extracolsep\fill}llllll@{\extracolsep\fill}}
    \toprule
    Zenith Angle & $I_{0}$       & $E_{0}$ & $n$     & $k$     & $\chi^{2}$/\rm{NDF} \\ 
    (degrees)    &  (m$^{-2}$s$^{-1}$sr$^{-1}$) & (GeV/g-cm$^{-2}$) &  & (GeV) &  \\
    \midrule
    9.07  & 69.770 {$\pm$ 5.768} & 0.005  & $3.28 \pm 0.00$ & 0.80 & {1.537} \\   
    16.96 & 67.445 {$\pm$ 7.412} & 0.005  & $3.28 \pm 0.01$ & 0.80 & {0.993} \\   
    23.22 & 64.601 {$\pm$ 6.688} & 0.005  & $3.28 \pm 0.01$ & 0.80 & {1.194} \\   
    30.40 & 59.513 {$\pm$ 6.444} & 0.005  & $3.29 \pm 0.01$ & 0.80 & {1.718} \\   
    38.05 & 50.436 {$\pm$ 6.033} & 0.005  & $3.29 \pm 0.01$ & 0.80 & {1.864} \\   
    44.56 & 39.285 {$\pm$ 4.352} & 0.005  & $3.26 \pm 0.01$ & 0.80 & {1.218} \\   
    50.39 & 31.453 {$\pm$ 4.191} & 0.005  & $3.25 \pm 0.01$ & 0.80 & {1.522} \\   
    55.77 & 23.075 {$\pm$ 3.968} & 0.005  & $3.22 \pm 0.02$ & 0.80 & {0.634} \\
    \bottomrule
  \end{tabular*}
  \label{tab:FinalFixZen}
\end{table*}

The set of fixed values for $E_0$ and $k$ that have been used in last subsection gives good
description of the muon flux distributions as a function of depth but do not reproduce
the data as a function of zenith angle variation.
Thus, we fixed the value of $k$ first to be 0.80 GeV and look for the variation in variable $E_0$.
A higher value of $E_0$ at 0.005 GeV/g-cm$^{-2}$ is required for zenith angle variation. 

Figure~\ref{fig:FinalFixDepZen} shows
the total muon flux at different zenith angles~\cite{ACHARD200415} at an altitude of
450 m. The solid lines are obtained by fitting the data with Eq.~\ref{eq:muonflux}
keeping $E_0$ = 0.005 GeV/g-cm$^{-2}$ and $k$ = 0.80 GeV.
The values of parameters of Eq. \ref{eq:muonflux} for different zenith angles
are given in Table~\ref{tab:FinalFixZen}.
At zenith angle 9.07$^{\circ}$, the integrated flux is 69.77 m$^{-2}$s$^{-1}$sr$^{-1}$.
The integrated flux decreases with increasing zenith angle which is expected as the
path length increases with zenith angle. The value of $n$ is almost constant with
variation in zenith angle.
{The parametrization provides a reasonable extrapolation in the 1–30 GeV/c range. Additional low-energy data would improve the description.}

\subsection{Variation of Integrated flux with respect to atmospheric depth and zenith angle} \label{subsec:3}

\begin{figure}[hbt!]
  \centering
  \includegraphics[width=8cm]{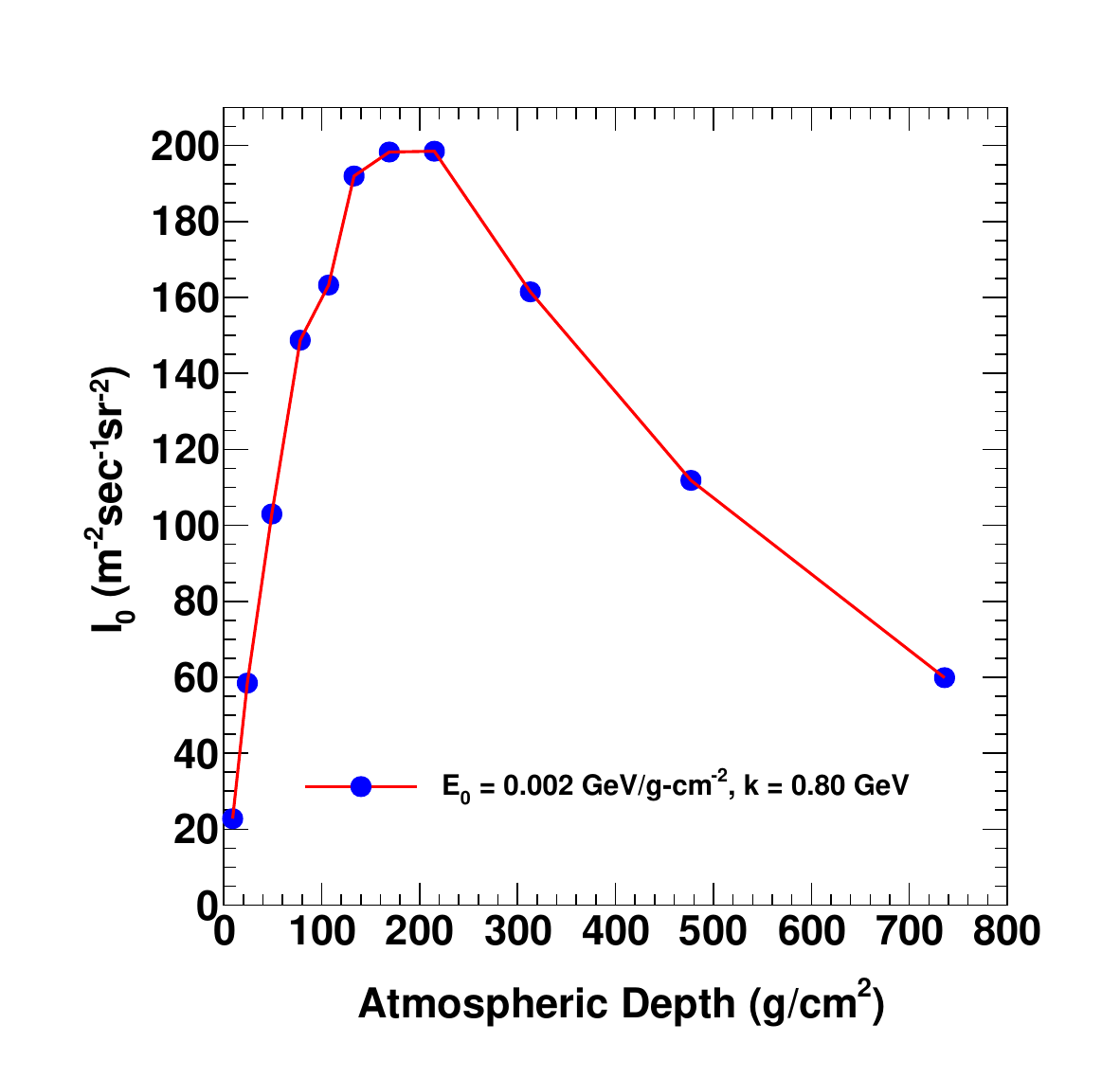}
  \includegraphics[width=8cm]{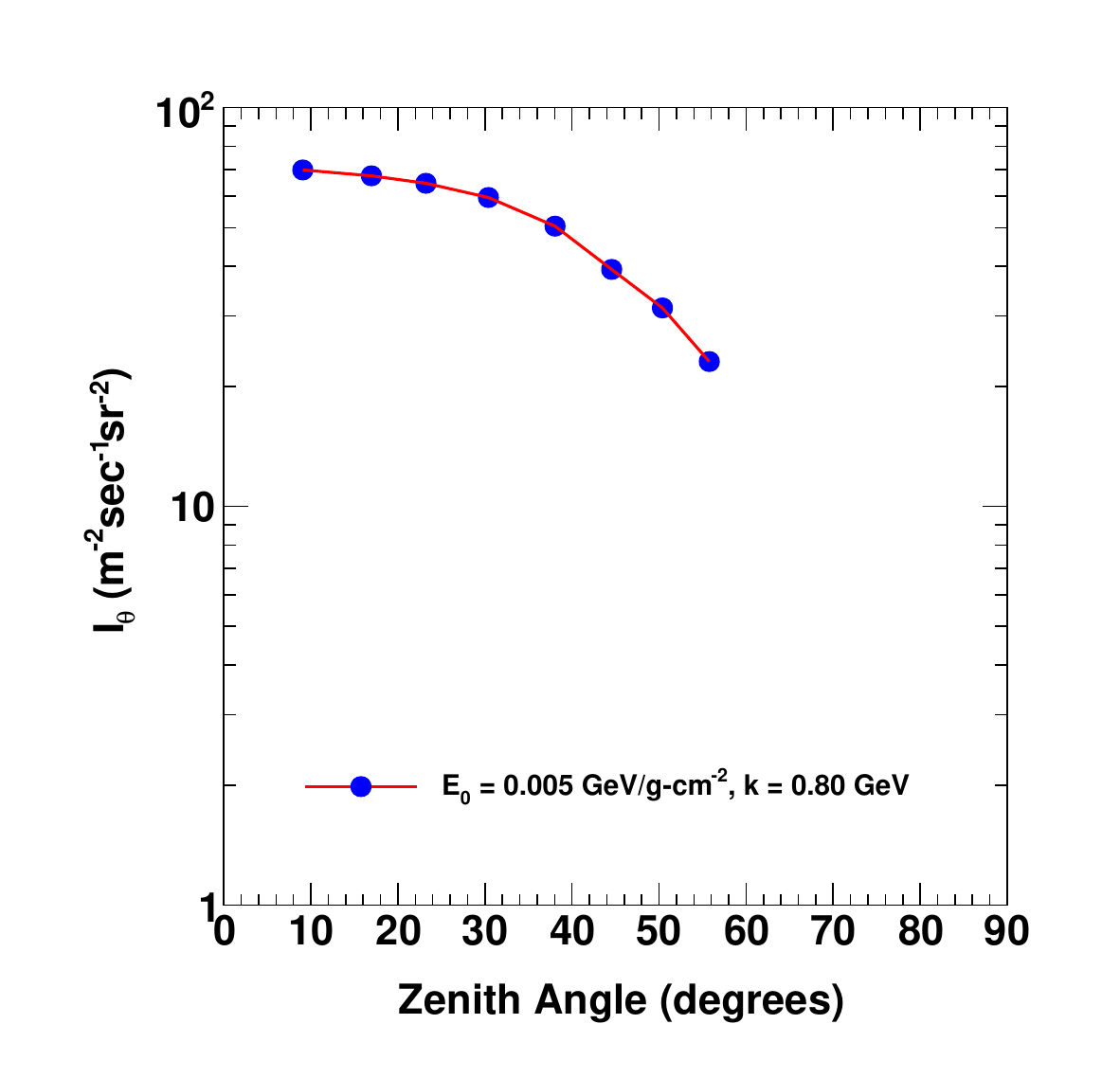}
  \caption{Variation in Integrated muon flux with respect to atmospheric depth (left) and zenith angle (right).}
  \label{fig:IntFluxVar}
\end{figure}

As discussed in the previous section, the parametrization of the data helps
to get the integrated flux as function of atmospheric depth and zenith angle. 
Figure~\ref{fig:IntFluxVar} (left) shows the variation in integrated negative muon flux with
respect to the atmospheric depth.
The shape of the curve can be understood as follows. 
As the primary particles enter the atmosphere they interact with the air molecules and this results
in the production of hadrons finally decaying to muons. As the particles travel downwards they
encounter more target nuclei and hence the production increases very fast. At a certain depth,
the muon flux reaches a maximum, beyond which it begins to decrease. This happens since the
produced muons also interact with the atmosphere and decay. Thus the maximum number
of muons are found at an atmospheric depth of about 200 gm/cm$^2$.

Figure~\ref{fig:IntFluxVar} (right) shows the variation in integrated total muon flux with respect to
the zenith angle.
The integrated muon flux decreases as the zenith angle increases
because muons arriving at larger zenith angles travel longer distances through the atmosphere,
which increases their probability of interacting and decaying, leading to a lower observed flux.

\section{The Distribution of Atmospheric Neutrinos} \label{sec:ParamNeut}
Along with muons and other secondary particles, neutrinos are also produced in the atmosphere during
extensive air showers created by primary particles.
However, their interaction cross-section is extremely low, allowing them to pass
through the atmosphere and Earth with {very few interactions}.  
Several big experimental detectors have been developed to measure their properties
in various laboratories, such as KamLAND, Gran Sasso,
IceCube etc.~\cite{HIRATA1988416,Barr:1989ru,PhysRevLett.69.1010,BECKERSZENDY1995331,Agrawal:1995gk,199833,PhysRevLett.113.101101,2017JInst..12P3012A,psf2023008062}
{Since the measured energy distributions of neutrinos are scarce,
  the present analysis employs neutrino flux data generated using the JAM
  (Jet AA Microscopic) nuclear interaction model, a widely used hadronic interaction
  framework that reproduces particle yields in cosmic-ray–induced air showers.
  This simulation takes measured data on primary particle distribution as input
  and uses known Physics processes.
  We specifically use the flux datasets provided by Honda et al.~\cite{PhysRevD.83.123001},
  who applied the JAM model to compute the atmospheric neutrino flux at several
  experimental locations worldwide.}

The neutrino flux is parametrized using the following modified energy distribution function:
\begin{equation}\label{eq:NeutFit}
  I(E) = I_{0}N\Bigr[\frac{E_{0} d}{\cos\theta} + E\Bigr]^{-n}\Bigr(1 + \frac{E}{\epsilon}\Bigr)^{-1}.
\end{equation}
Unlike other secondary particles discussed in the previous sections, such as muons,
neutrinos do not decay in the atmosphere. As a result, the decay term has been excluded
from the distribution. Instead, we introduce a source term, which accounts for an additional
inverse dependence of the neutrino flux on energy, beyond the standard power-law behavior.  
This term becomes significant for neutrinos with energies exceeding a certain threshold,
governed by the parameter $\epsilon$, which effectively encodes the influence of the finite
lifetimes of the parent particles that produce the neutrinos.

\begin{figure}[hbt!]
  \centering
  \includegraphics[width=8cm]{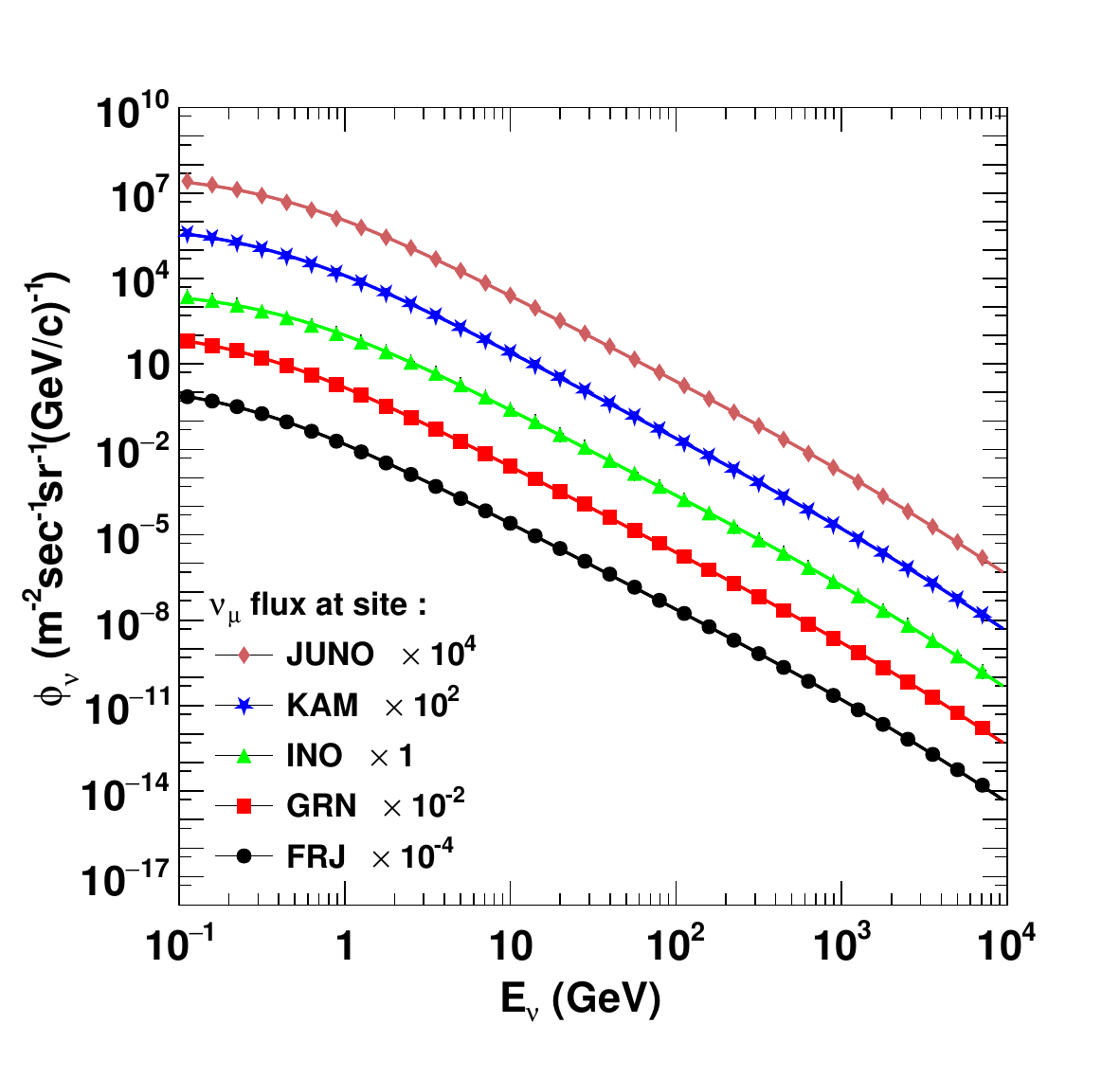}
  \includegraphics[width=8cm]{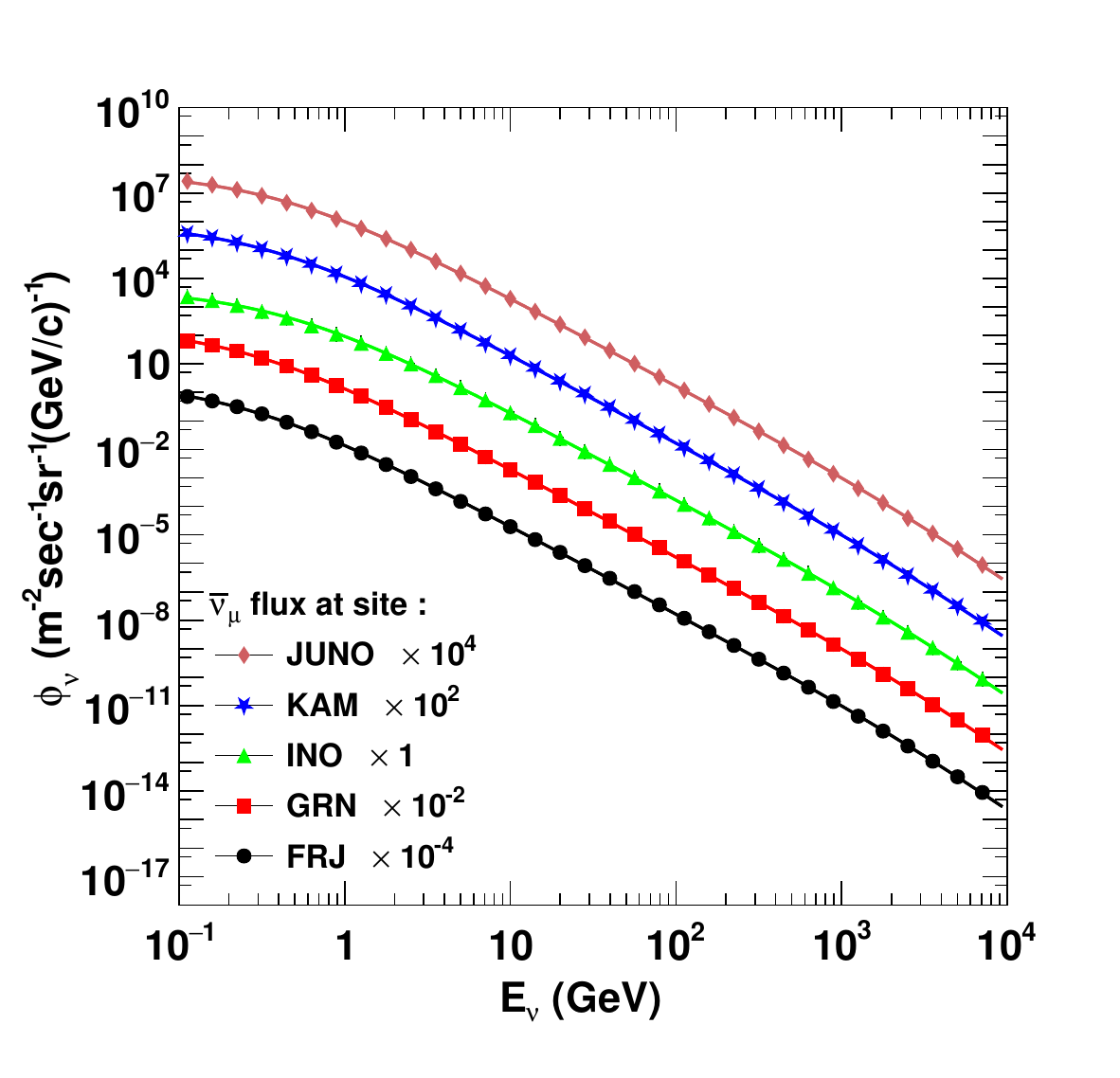}
  \caption{The energy distribution of muon type atmospheric neutrinos (left) and anti-neutrinos (right),
    simulated for various sites~\cite{PhysRevD.83.123001} for $\cos\theta$ = 0.90 - 1.00, the
    solid lines are fits using Eq.~\ref{eq:NeutFit}.}
  \label{fig:NeuMu}
\end{figure}

\begin{figure}[hbt!]
  \centering
  \includegraphics[width=8cm]{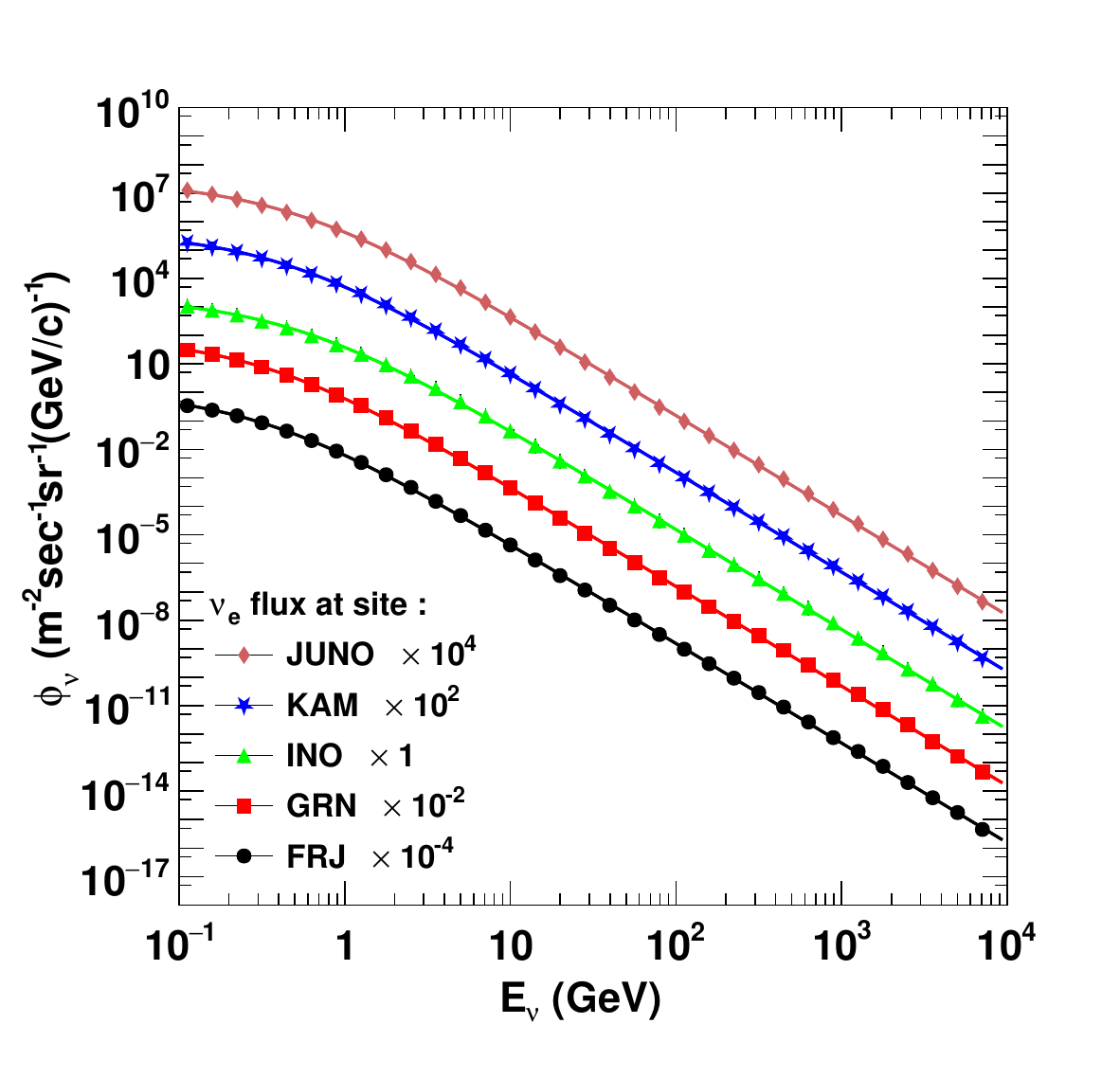}
  \includegraphics[width=8cm]{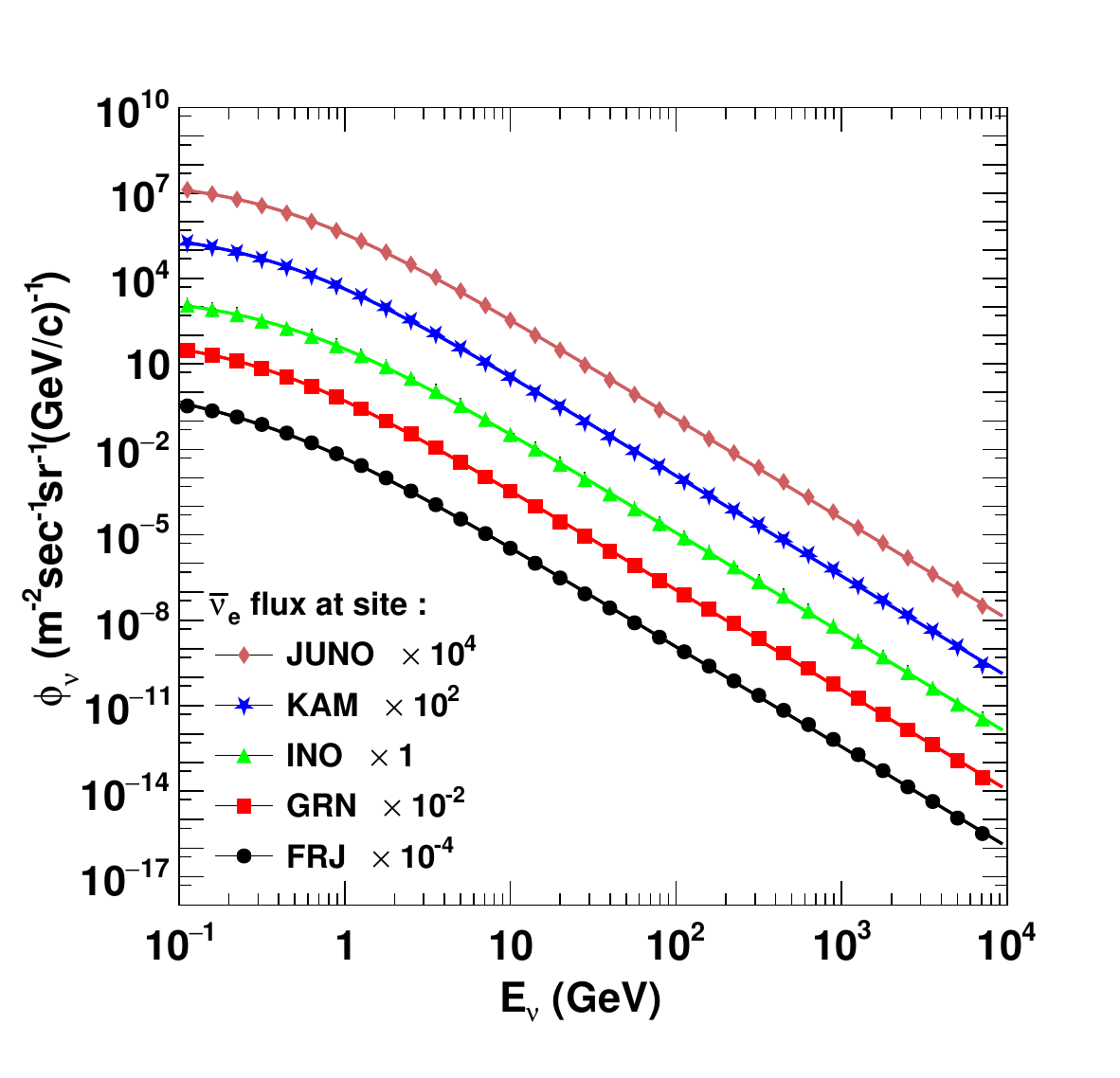}
  \caption{The energy distribution of electron type atmospheric neutrinos (left) and anti-neutrinos (right),
    simulated for various sites~\cite{PhysRevD.83.123001} for $\cos\theta$ = 0.90 - 1.00, the
    solid lines are fits using Eq.~\ref{eq:NeutFit}.}
  \label{fig:NeuElec}
\end{figure}

Figure~\ref{fig:NeuMu} shows the energy distribution of muon type atmospheric neutrinos (left) and anti-neutrinos (right),
simulated for various sites~\cite{PhysRevD.83.123001} for $\cos\theta$ = 0.90 - 1.00, the
solid lines are fits using Eq.~\ref{eq:NeutFit}.
Figure~\ref{fig:NeuElec} shows the same for electron-type neutrinos (and anti-neutrinos). 
The corresponding fit parameters are listed in Tables~\ref{tab:NeuMu} and~\ref{tab:NeuElec}.
The function in Eq.~\ref{eq:NeutFit} gives excellent description of all types
of neutrinos.

\begin{table*}[hbt!]
  \centering
  \caption{Values of parameters obtained by fitting the simulated data for the muon type atmospheric
    neutrinos and anti-neutrinos at various sites for $\cos\theta$ = 0.90 - 1.00 with
    Eq.~\ref{eq:NeutFit}.}
  \begin{tabular*}{\textwidth}{@{\extracolsep\fill}lllllll@{\extracolsep\fill}}
    \toprule
    Site & Particle & $I_{0}$ & $E_{0}.d$       & $n$     & $\epsilon$  & $\chi^{2}$/\rm{NDF} \\ 
    & Type   &  (m$^{-2}$s$^{-1}$sr$^{-1}$) & (GeV) &  & (GeV) &  \\
    \midrule
    Frejus  & $\nu_{\mu} $ & 1312.47 {$\pm$ 39.57} & 0.22 $\pm$ 0.01  & 3.02 $\pm$ 0.01 & 2123.34 $\pm$ 195.28 & 0.46 \\ 
    (45.18$^{\circ}$N, 6.69$^{\circ}$E)& $\bar{\nu}_{\mu}$ & 1291.69 {$\pm$ 37.95} & 0.21 $\pm$ 0.01 & 3.09 $\pm$ 0.01 & 2351.15 $\pm$ 217.68 & 0.39 \\  
    Gran Sasso & $\nu_{\mu} $ & 1209.35 {$\pm$ 37.18} & 0.24 $\pm$ 0.01 & 3.03 $\pm$ 0.01 & 2183.32 $\pm$ 203.32 & 0.45\\ 
    (42.42$^{\circ}$N, 13.51$^{\circ}$E)& $\bar{\nu}_{\mu}$ & 1188.68 {$\pm$ 35.57} & 0.23 $\pm$ 0.01 & 3.09 $\pm$ 0.01 & 2403.84 $\pm$ 226.01 & 0.42\\  
    Kamioka & $\nu_{\mu} $ &  829.43 {$\pm$ 27.76} & 0.31 $\pm$ 0.01  & 3.04 $\pm$ 0.01 & 2276.77 $\pm$ 217.44 & 0.46 \\ 
    (36.42$^{\circ}$N, 137.31$^{\circ}$E)& $\bar{\nu}_{\mu}$ &  811.19 {$\pm$ 26.43} & 0.30 $\pm$ 0.01 & 3.10 $\pm$ 0.01 & 2453.71 $\pm$ 237.91 & 0.41 \\  
    JUNO  & $\nu_{\mu} $ &  617.34 {$\pm$ 22.41} & 0.37 $\pm$ 0.01 & 3.03 $\pm$ 0.01 & 2069.50 $\pm$ 200.97 & 0.95 \\ 
    (22.11$^{\circ}$N, 112.52$^{\circ}$E)& $\bar{\nu}_{\mu}$ &  600.24 {$\pm$ 20.98} & 0.36 $\pm$ 0.01 & 3.09 $\pm$ 0.01 & 2370.01 $\pm$ 229.01 & 0.58 \\ 
    INO   & $\nu_{\mu} $ & 550.46 {$\pm$ 20.68} & 0.40 $\pm$ 0.01 & 3.03 $\pm$ 0.01  & 2083.80 $\pm$ 260.79 & 1.36 \\ 
    (9.96$^{\circ}$N, 77.28$^{\circ}$E)& $\bar{\nu}_{\mu}$ & 532.91 {$\pm$ 19.24} & 0.38 $\pm$ 0.01 & 3.09 $\pm$ 0.01  & 2339.75 $\pm$ 232.85 & 0.88 \\
    \bottomrule
  \end{tabular*}
  \label{tab:NeuMu}
\end{table*}

\begin{table*}[hbt!]
  \centering
  \caption{Values of parameters obtained by fitting the simulated data for the electron type
    atmospheric neutrinos and anti-neutrinos at various sites for $\cos\theta$ = 0.90 - 1.00
    with Eq.~\ref{eq:NeutFit}.}
  \begin{tabular*}{\textwidth}{@{\extracolsep\fill}lllllll@{\extracolsep\fill}}
    \toprule
    Site & Particle  & $I_{0}$ &  $E_{0}.d$ &     $n$     & $\epsilon$  & $\chi^{2}$/\rm{NDF} \\  
    & Type  &  (m$^{-2}$s$^{-1}$sr$^{-1}$) & (GeV) &  & (GeV) &  \\  
    \midrule
    Frejus     & $\nu_{e} $ & 625.73 {$\pm$ 17.73} & 0.28 $\pm$ 0.02 & 2.50 $\pm$ 0.00 & 0.29 $\pm$ 0.06 & 2.20\\
    (45.18$^{\circ}$N, 6.69$^{\circ}$E)& $\bar{\nu}_{e}$ & 561.53 {$\pm$ 13.39} & 0.43 $\pm$ 0.01 & 2.52 $\pm$ 0.00 & 0.03 $\pm$ 0.00 & 3.38 \\  
    Gran Sasso & $\nu_{e} $ & 570.84 {$\pm$ 17.55} & 0.30 $\pm$ 0.04 & 2.50 $\pm$ 0.00 & 0.31 $\pm$ 0.09 & 2.40 \\ 
    (42.42$^{\circ}$N, 13.51$^{\circ}$E)& $\bar{\nu}_{e}$ & 516.68 {$\pm$ 26.19} & 0.33 $\pm$ 0.01  & 2.51 $\pm$ 0.00 & 0.19 $\pm$ 0.13 & 2.83 \\  
    Kamioka    & $\nu_{e}  $ &  378.94 {$\pm$ 8.85} & 0.37 $\pm$ 0.06  & 2.50 $\pm$ 0.00 & 0.39 $\pm$ 0.16 & 2.22 \\ 
    (36.42$^{\circ}$N, 137.31$^{\circ}$E)& $\bar{\nu}_{e}$ &  355.30 {$\pm$ 8.30} & 0.47 $\pm$ 0.00 & 2.52 $\pm$ 0.00 & 0.16 $\pm$ 0.01 & 2.92 \\  
    JUNO       & $\nu_{e}  $ &  273.90 {$\pm$ 6.53} & 0.35 $\pm$ 0.04 & 2.51 $\pm$ 0.01 & 0.80 $\pm$ 0.20 & 2.90 \\ 
    (22.11$^{\circ}$N, 112.52$^{\circ}$E)& $\bar{\nu}_{e}$ &  265.90 {$\pm$ 3.98} & 0.29 $\pm$ 0.02 & 2.52 $\pm$ 0.01 & 0.87 $\pm$ 0.17 & 2.58 \\  
    INO        & $\nu_{e}  $ & 239.50 {$\pm$ 3.84} & 0.34 $\pm$ 0.03  & 2.51 $\pm$ 0.01  & 1.01 $\pm$ 0.18 & 3.13 \\ 
    (9.96$^{\circ}$N, 77.28$^{\circ}$E)& $\bar{\nu}_{e}$ & 238.22 {$\pm$ 3.44} & 0.29 $\pm$ 0.02 & 2.52 $\pm$ 0.01  & 1.02 $\pm$ 0.17 & 2.44 \\  
    \bottomrule
  \end{tabular*}
  \label{tab:NeuElec}
\end{table*}

{Muon-type neutrinos are mainly produced in the atmosphere
  through the decay of charged pions and kaons, which generate muons and their
  associated neutrinos. At higher energies, these mesons have longer lifetimes
  due to relativistic time dilation and are more likely to interact before decaying,
  reducing the number of neutrinos produced.
  Thus the large value of parameter $\epsilon$ gives the energy above which the
  spectrum starts becoming steeper as effectively it makes the spectral index
  to increase.
  In contrast, electron-type neutrinos originate primarily from muon decay.
  Since muons live much longer than their parent mesons, this transition starts at
  much lower energy giving small $\epsilon$.}
The parameter $I_0$ for neutrinos is slightly higher than that for anti-neutrinos
at all sites. $I_0$ for muon-type neutrinos is more than twice of that for
electron-type neutrinos.
$E_0d$ is very low for all types of neutrinos.
The index $n$ is around 3.00 for muon-type neutrinos, which closely matches the
$n$ value for muons and 2.50 for electron-type neutrinos.
An analytic function with parameters at different sites can be very useful
for detector simulation, saving computational time.
{The function given by Eq.~\ref{eq:NeutFit} works well for the
  energy distribution of neutrinos in all energy range. The data was taken from
  the simulation at various sites as the actual experimental data for neutrino
  energy distribution are scarce. The simulated data sets  of neutrinos are obtained
  using measured primary particle distributions as inputs and detailed simulation
  including known processes in the atmosphere. These simulated data sets are very
  well represented by our analytic function which can be useful for detector simulation.
  In future, measured energy distributions of atmospheric neutrinos
  (e.g., from Super-Kamiokande) may become available which will useful to validate
  the simulation/parametrizations. }

\section{Summary and Prospects} \label{sec:Summ}
In this work, we presented analytical formulations to describe the momentum distribution
of primary and secondary cosmogenic particles.
Although distributions of different particles are described by power law, their
modification terms are different depending on their interaction.
We begin by modifying the power-law distribution and incorporating terms that account
for energy loss at lower energies for both primary and secondary particles.  
For primary particles such as protons and helium, a $1/E$ term is introduced
to account for flux loss.  
For secondary particles such as muons, an exponentially decaying factor is included
to reflect their finite lifetime.  
Additionally, these modified formulations incorporate parameters such as $d$ (atmospheric depth)
and $\theta$ (zenith angle) to account for variations in flux with altitude and zenith angle.  
It is shown that the presence of decay term does not affect the zenith angle distribution
of muons. We fit various experimental data at sea level as well as at different
altitudes and zenith angles in terms of parameters of the proposed analytical functions. 
Furthermore, we extend our parametrization to atmospheric muon and electron-type neutrinos
by introducing a source term instead of a decay function which are applied on simulated data.
The proposed functions successfully describe the momentum distributions of
primary cosmic particles as well as secondary particles such as muons and neutrinos. 
The functions are useful for obtaining integrated flux for different cases and as inputs
for detector simulations.
{These formulations can be extended to other cosmogenic particles,
  such as charm-induced leptons (prompt muons and neutrinos) from decays of $D$-mesons
  and $\Lambda_c$-baryons, and to high-energy photons from neutral pion decay.
  Similar analytical treatments may also apply to heavier primary nuclei beyond helium.
  The present parametrizations are valid within the measured energy range and effects
  like geomagnetic cutoffs etc. are absorbed in the fitted parameters.} 


\bmsection*{Financial disclosure}

No financial support has been used for this work

\bmsection*{Conflict of interest}

The authors declare no potential conflict of interests.

\bibliography{wileyNJD-AMA}

\end{document}